\documentclass[12pt,a4paper]{article}
\usepackage{fullpage}
\usepackage{epsfig,multicol,bbm}
\usepackage{axodraw}
\usepackage{multirow}
\newcommand\fverbdo{\egroup\medskip\noindent%
            \fbox{\unhbox\fverbbox}\ }
\newcommand\fverbit{\egroup\item[\fbox{\unhbox\fverbbox}]}
\newbox\fverbbox

\begin{document}
\begin{titlepage}







\begin{center}
\large\bfseries
R2SM: a package for the analytic computation of the ${\bf R_2}$ Rational
terms in the Standard Model of the Electroweak interactions
\\[0.5cm]
\normalsize\normalfont
M.V. Garzelli$^{1}$ and I. Malamos$^2$
\\[0.3cm]
\small\itshape
$^1$INFN, Italia \&
Departamento de F\'{i}sica Te\'orica y del Cosmos y CAFPE, 
\\
\small\itshape
Universidad de Granada, E--18071 Granada, Spain
\\[0.2cm]
$^2$Department of Theoretical High Energy Physics,
        Institute for Mathematics, Astrophysics and Particle Physics, 
        Radboud Universiteit Nijmegen,  
        6525 AJ Nijmegen, the Netherlands. 
\\[0.3cm]
\small\itshape
e-mail: {\rm\texttt{garzelli@to.infn.it}}, {\rm\texttt{J.Malamos@science.ru.nl}}
\end{center}

\begin{abstract}
The analytical package written in FORM presented in this paper allows 
the computation of the complete set of Feynman Rules producing the 
Rational terms of kind ${\rm R_2}$ contributing to the virtual part of NLO 
amplitudes in the Standard Model of the Electroweak  
interactions. 
Building block 
topologies filled 
by means of
generic scalars, vectors and fermions,
allowing to build these Feynman Rules in terms of specific 
elementary particles,
are explicitly given in the $R_\xi$ gauge class, together with the automatic 
dressing procedure to obtain the Feynman Rules from them. The results 
in more specific gauges, like the 't Hooft Feynman one,  
follow as particular cases,  
in both the HV and the FDH dimensional regularization schemes.
As a check on our formulas, 
the gauge independence of the total Rational contribution 
(${\rm R_1}$ + ${\rm R_2}$)
to renormalized S-matrix elements is verified by considering the specific 
example of the 
H~$\rightarrow$~$\gamma \gamma$ decay process at 1-loop.  
This package can be of interest for people aiming at a better
understanding of the nature of the Rational terms. It is organized
in a modular way, allowing a further use of some its files  
even in different contexts. Furthermore, it can be considered as 
a first seed in the effort towards a complete automation of the 
process of the analytical calculation of the ${\rm R_2}$ effective vertices, 
given the Lagrangian of a generic 
gauge theory of 
particle interactions.

\end{abstract}

\nonumber

\end{titlepage}

\newcounter{im}
\setcounter{im}{0}
\newcommand{\exampleSp}{\stepcounter{im}\includegraphics[scale=0.9]{SpinorExamples_\arabic{im}.eps}}
\newcommand{\myindex}[1]{\label{com:#1}\index{{\tt #1} & pageref{com:#1}}}
\renewcommand{\topfraction}{1.0}
\renewcommand{\bottomfraction}{1.0}
\renewcommand{\textfraction}{0.0}
\newcommand{\nn}{\nonumber \\}
\newcommand{\eqn}[1]{eq.~\ref{eq:#1}}
\newcommand{\be}{\begin{equation}}
\newcommand{\ee}{\end{equation}}
\newcommand{\ba}{\begin{array}}
\newcommand{\ea}{\end{array}}
\newcommand{\bea}{\begin{eqnarray}}
\newcommand{\eea}{\end{eqnarray}}
\newcommand{\bqa}{\begin{eqnarray}}
\newcommand{\eqa}{\end{eqnarray}}
\newcommand{\nl}{\nonumber \\}
\def\db#1{\bar D_{#1}}
\def\zb#1{\bar Z_{#1}}
\def\d#1{D_{#1}}
\def\tld#1{\tilde {#1}}
\def\slh#1{\rlap / {#1}}
\def\eqn#1{eq.~(\ref{#1})}
\def\eqns#1#2{Eqs.~(\ref{#1}) and~(\ref{#2})}
\def\eqnss#1#2{Eqs.~(\ref{#1})-(\ref{#2})}
\def\fig#1{Fig.~{\ref{#1}}}
\def\figs#1#2{Figs.~\ref{#1} and~\ref{#2}}
\def\sec#1{Section~{\ref{#1}}}
\def\app#1{Appendix~\ref{#1}}
\def\tab#1{Table~\ref{#1}}
\def\cg{c_\Gamma}
\newcommand{\bfig}{\begin{center}\begin{picture}}
\newcommand{\efig}[1]{\end{picture}\\{\small #1}\end{center}}
\newcommand{\flin}[2]{\ArrowLine(#1)(#2)}
\newcommand{\ghlin}[2]{\DashArrowLine(#1)(#2){5}}
\newcommand{\wlin}[2]{\DashLine(#1)(#2){2.5}}
\newcommand{\zlin}[2]{\DashLine(#1)(#2){5}}
\newcommand{\glin}[3]{\Photon(#1)(#2){2}{#3}}
\newcommand{\gluon}[3]{\Gluon(#1)(#2){5}{#3}}
\newcommand{\lin}[2]{\Line(#1)(#2)}
\newcommand{\sof}{\SetOffset}

\section{Introduction}
\label{intro}

The calculation of NLO radiative corrections to multiparticle production
processes
has achieved significant progresses in the last few years, thanks 
to refinements to the traditional techniques~\cite{pasve,dennerred,golem} 
on the one hand, and to 
the introduction of new methods, mainly based on Unitarity~\cite{uni0,uni}
and Generalized Unitarity~\cite{genuni} principles, on the other. 
This has already allowed the computation of several (differential) 
cross-sections at NLO, especially for key signal and background 
particle scatterings and decays of interest at colliders, 
whose signatures can potentially 
through light on 
the mechanism underlying 
the Electroweak (EW) Symmetry Breaking process~\cite{leshouches,nlopapers}. 

A generic 1-loop amplitude can be decomposed as a linear combination of
known scalar integrals, with up to 4 external legs, plus a residual 
Rational part R. 
In the framework of the approaches based on the Unitarity
of the S-matrix, the first piece is Cut-Constructible (CC),
i.e. it can be reconstructed by properly cutting the loop 
amplitude in tree-level like sub-amplitudes, thus conceptually 
reducing the complexity of the calculation of 1~-~loop integrals
to a simpler computation of tree-level diagrams. 
On the other hand, when working in approaches 
based on 4 integer dimensions, the R terms, or at least some
of them as clarified in the following,  
cannot indeed be calculated just in terms of 
tree-level diagrams, but a full 1-loop computation is 
needed.
This can be performed according to the
traditional Feynman diagram approach~\cite{yang, binothrational}. 
An alternative strategy consists in making use of Unitarity-based 
on-shell recursion relations~\cite{carolaonshell}, 
leading to a recursive evaluation 
of R by means of bootstrapping techniques~\cite{boot}. 
On the other hand, it is worth observing that the R terms can be put 
on the same footing as the CC part at the price 
of introducing a larger number of integer dimensions,
as worked out in the $d$-dimensional extensions of the
Unitarity approach~\cite{ddimuni}.
Despite the attractive elegance of the $d$-dimensional Unitarity formulations,
in this paper we choose to follow a variation of 
the first approach, thus avoiding to introduce
extra integer dimensions (and, as a consequence, 
to extend to these higher dimensions the external particle wave functions),
and just allowing $d = 4 + \epsilon$ dimensions 
in the dimensional regularization scheme we use to regularize the
divergencies appearing during the computation of 1-loop integrals
(even in case of finite tensor integrals, singularities may  
arise in the tensor reduction process and need to be regularized). 
We worked in the framework of the OPP method~\cite{opp}, 
one of the Unitarity 
inspired algebraic and universal (i.e. independent from the model
of particle interactions at hand) procedures first introduced to 
automatically calculate the CC part of any 1-loop amplitude. 
In the OPP approach, the R terms can be organized 
in two classes, according to their origin: the ${\rm R_1}$ terms, 
arising from the
mismatch between the 4-dimensional part of the numerator 
and the $d$-dimensional denominator including the 
poles
of 
the propagators in the regularized integrand of 1-loop integrals, 
and the ${\rm R_2}$ terms, arising instead from the $\epsilon$-dimensional
part of the numerator. Both these classes are thus a residual
effect of the dimensional regularization scheme one introduces
on the integrands to calculate 1-loop integrals.
In the framework of OPP, it has been shown that the ${\rm R_1}$ terms
are closely related to the CC part, and can indeed be obtained
numerically {\it at the same time} of the last one~\cite{rational}. 
Unfortunately, however, this property does not apply to
the ${\rm R_2}$ terms, that, instead, need a dedicated computation. 
Anyway, the problem of the calculation of the ${\rm R_2}$ part of a generic
1-loop amplitude can be solved by observing that only 1-particle
irreducible diagrams with up to 4 external legs can contribute 
to ${\rm R_2}$, due to the exclusively ultraviolet nature of these terms, 
as proven in~\cite{binothrational}.
In other words, ${\rm R_2}$ terms will never contribute to the 
$1/\epsilon$ and $1/\epsilon^2$ pole parts of 1-loop 
amplitudes, whose nature is instead completely infrared. 
Thus, taking into account that the number of the contributing diagrams is
limited, it is possible to calculate, once and for all for the theory
of interaction at hand, all possible ${\rm R_2}$ effective vertices, up to 
4-external legs, and then use these effective vertices as building blocks 
for computing the ${\rm R_2}$ contribution to each specific 1-loop 
amplitude~\cite{rational}. 
Given the set of all external particles (with their momenta and their
quantum numbers) identifying a 1-loop helicity amplitude, it will be enough 
to consider  all tree-level diagrams joining them, 
which include one (and only one) ${\rm R_2}$ effective vertex.  
This recipe has been adopted in the HELAC-1-loop code~\cite{proof},
where the ${\rm R_2}$ effective vertices entering QCD 1-loop 
corrections~\cite{qcdrational} have been implemented, and recursion 
relations have been adopted to evaluate the corresponding contributions 
to the amplitudes.
Furthermore, our ${\rm R_2}$ Feynman Rules, together with 
(off-shell) recursive relations for tensor integrals,
have been applied to the numerical computation of the ${\rm R_2}$ 
contribution to 1-loop multigluon amplitudes
in a tensor reduction framework~\cite{andre}.

Throughought this paper we denote by {\it topology} 
a set of lines (propagators) connecting a set of points (vertices),
by {\it generic diagram} a topology filled by means of generic
Standard Model (SM)
scalar $s$, vector $v$ and/or fermion $f$ fields, and by
{\it particle diagram} a generic diagram whose components have
been further specified in terms of selected SM
particles (like $H$, $Z$, $e^-$, etc.).
After a formal definition of the R terms in Section~\ref{section2}
and our comments on the gauge choice in Section~\ref{section3},
we present our analytical results for the ${\rm R_2}$ contributions 
to all leading generic 1-loop 1-particle irreducible diagrams with up to 
4 external legs arising in the SM of EW interactions, diagram by
diagram, 
and the FORM code we have written to obtain from 
them the ${\rm R_2}$ effective vertices involving real specific particles, to be
used in actual calculation of 1-loop amplitudes, in
Section~\ref{program}. These analytical 
results are of interest for people aiming at better understanding 
the nature and the properties of the ${\rm R_2}$ terms.
We include a test of the reliability of our code together with 
considerations concerning the gauge invariance of R in Section~\ref{hgg}, 
and we draw our conclusions in Section~\ref{conclusions}. 
Additional information about the notation used in our code 
is available in the Appendix 
and in the README file within the package.\\

\section{The R contribution to 1-loop amplitudes}
\label{section2}

In this Section we formally define the R terms, and in particular 
the ${\rm R_2}$ class, briefly reviewing how they appear 
in the computation of 1-loop amplitudes. 
Our starting point is the $integrand$
of a generic $m$-point amplitude, written as
\begin{eqnarray}
A(q) = \frac{N(q)}{D_0 D_1....D_{m-1}} \, , \,\,\, {\mathrm{with}} \,\,\, D_i=(q+p_i)^2 - m_i^2 \, , 
\end{eqnarray}
where $q$ is the loop momentum, $q+p_i$ and $m_i$ are the momentum and the mass
of the $i$-th loop particle  ($i$ = 0, 1,....$m-1$). 

A dimensional regularization procedure is then introduced, at the aim of
regularizing the 
singularities
in the integrand amplitude, in order to be allowed 
to evaluate its integral. 
In renormalizable gauge theories, 
the number of ultraviolet (UV) divergent integrals is finite, 
and the UV divergencies are re-absorbed, after integral calculation, 
in renormalized quantities. The infrared (IR) singularities, instead, cancel 
when combining together the real and the virtual contribution 
to the amplitude at each fixed order in the perturbative expansion in terms
of the coupling constant, according to the KLN theorem~\cite{kln}. 
Thus the results
for 1-loop amplitudes will, in general, include a finite part plus a divergent
part, of IR origin only. 
  
We choose to work in $d=4+\epsilon$ dimensions (dim).
The extension to $d$-dim of the
integrand of the amplitude, $N(q)$ $\rightarrow$ $\bar{N}(\bar{q})$, 
is achieved through the transformations:
\begin{eqnarray}
q_\mu  \rightarrow \bar{q}_{\mu} & = & q_\mu + \tilde{q}_{\tilde{\mu}} \, ,\\
\gamma_\mu \rightarrow \bar{\gamma}_{\bar{\mu}} & = & \gamma_\mu + \tilde{\gamma}_{\tilde{\mu}} \, ,\\
g_{\mu\nu} \rightarrow \bar{g}_{\bar{\mu}\bar{\nu}} & = & g_{\mu\nu} + \tilde{g}_{\tilde{\mu}\tilde{\nu}} \, ,    
\end{eqnarray}
where we have denoted quantities in $d$-dim with a bar, and we have explicitly shown the $\epsilon$-dim part of each quantity, including tildes. 
The quantities in $\epsilon$-dim are orthogonal with
respect to the quantities in 4-dim, thus, in developing the expression of $\bar{N}(\bar{q})$, one has to take into account relations like
\begin{eqnarray}
\bar{q}_{\bar{\mu}} v^\mu & = & q_\mu v^\mu \, , \\
\tilde{g}_{\tilde{\mu}\tilde{\nu}} {g}^{{\mu}{\nu}} &=& 0 
\, ,
\end{eqnarray}
and similar ones. 

One can thus rewrite the numerator in $d$-dim as a sum of a 4-dim part plus 
a residual part. 
The ${\rm R_2}$ terms correspond to the integral of the 
$\epsilon$-dim part of $\bar{N}(\bar{q})$, i.e.
\begin{eqnarray}
{\rm R_2} = lim_{\epsilon \rightarrow 0} \frac{i}{16 \pi^2} 
\int d^d \bar{q} \frac{\tilde{N}(q,\tilde{q}^2, \epsilon)}{\bar{D}_0 \bar{D}_1
\bar{D}_2.......\bar{D}_{m-1}} \, .
\label{eqr2}
\end{eqnarray}

To understand the origin of the CC and the ${\rm R_1}$ terms, one can instead expand
the 4-dim part of the numerator, $N(q)$, in terms of 4-dim
denominators $D_i$, according to the universal OPP decomposition~\cite{opp}, 
and then observe that $d$-dim denominators instead 
of 4-dim denominators appear in the dimensionally regularized
integrand of the amplitude~\cite{sixphoton}.
By rewriting these $d$-dim denominators 
in terms of the 4-dim denominators of the decomposition according to 
\begin{eqnarray}
\frac{1}{\bar{D}_i} =\frac{D_i}{\bar{D}_i D_i} = 
\left(1 - \frac{\tilde{q}^2}{\bar{D}_i}\right)\frac{1}{D_i} \, ,
\end{eqnarray}
one can see that the first part of the expression in the parentheses 
lead to the CC part of the amplitude (all terms in 4-dim,
as tree-level diagrams), whereas the second part of the expression
in the parantheses causes the appearence of the ${\rm R_1}$ contribution
to the amplitude:
\begin{eqnarray}
R_1 = lim_{\epsilon \rightarrow 0} \frac{i}{16 \pi^2}
\int d^d q \frac{f(q,\tilde{q}^2)}{\bar{D}_0 \bar{D}_1
\bar{D}_2.......\bar{D}_{m-1}} \, .
\end{eqnarray}

\section{Choice of a gauge}
\label{section3}

One of the crucial points to actually perform the computation of the ${\rm R_2}$
terms is the choice of a convenient gauge. 
We work in the $R_\xi$ class of gauges. 
More precisely, we consider the generalized
$R_\xi$ gauges, characterized by a three-parameter $\xi_A$, $\xi_Z$,
$\xi$ gauge fixing term. The standard $R_\xi$ gauges can be obtained
as the particular case corresponding to $\xi_A$ = $\xi_Z$ = $\xi$.

The main difference between
the gauges in this class and the unitary one, 
is the appearence in the first ones of unphysical scalars and 
Fadeev-Popov-DeWitt ghosts, as loop particles. 
The unitary (even called physical) gauge instead, comes out as the particular 
limit $\xi,\, \xi_Z \rightarrow \infty$ and $\xi_A \rightarrow $ finite
number
(these two independent limits are possible
because the first one does not fix the electromagnetic gauge invariance), 
corresponding to
unphysical particles becoming extremely heavy and decoupling
from the theory.\footnote{ 
Actually, the Fadeev-Popov ghosts do not completely decouple from the
theory. As 
$\xi, \, \xi_Z \rightarrow \infty$,
there are some surviving pieces (Lee-Yang terms) 
coming from the H-ghost-ghost vertices proportional 
to $\xi, \,\xi_Z$~\cite{lee}. 
} 
In fact, the unitary gauge condition
eliminates 3 of the 4 degrees of freedom in the scalar Higgs doublet,
re-emerging as longitudinal spin states of the $Z$ and $W^\pm$ 
gauge bosons acquiring a mass.
Thus, one has to deal with a reduced number of 
particles, i.e. with the physical ones only.  
This reduced number of particles is 
the reason why, when working at tree-level, unitary gauge is often preferred. 
However,
although
loop calculations
can indeed be performed even in this gauge, 
this is not convenient, due to the fact that the unitary gauge
is not manifestly renormalizable. In particular, 
the UV behavior of the theory is not properly manifest  and
appears worse than it really is. The reason is that the expression of
the propagators of the massive gauge bosons in the unitary gauge 
\begin{eqnarray}
\frac{ - i }{q^2-m_i^2} \left(g_{\mu\nu} - \frac{q_\mu q_\nu}{m_i^2}\right)
\label{propuni} \,\,\,{\rm with} \,\,\, i = W,\, Z
\end{eqnarray}
does not fall to zero for large momenta $q \rightarrow \infty$.

On the other hand, $R_\xi$ gauges are covariant
gauges, where the unphysical degrees of freedom in the Higgs
doublet do not disappear. However, the gauge fixing part of the 
Lagrangian eliminates the possibility of their mixing with gauge bosons,
thus allowing the development of a perturbation theory, 
and assigns them a mass which is
related to the one of the gauge bosons by the conditions
$m_\chi = \sqrt{\xi_Z} m_Z$, $m_\phi = \sqrt{\xi} m_W$.   
The propagators of the vector bosons in the generalized $R_\xi$ gauges have 
the expression
\begin{eqnarray}
\frac{ - i }{q^2-m_i^2} 
\left(g_{\mu\nu} + (\xi_i -1)\frac{q_\mu q_\nu}{q^2 - \xi_i m_i^2}\right)
\,\,\, \mathrm{with} \,\,\, \xi_i = \xi,\, \xi_Z, \,\xi_A \,\,\,\, (i = W, Z, A)
\label{proprxi}
\end{eqnarray}
 that, in the limit $\xi, \, \xi_{Z} \rightarrow \infty$, tends to eq.~(\ref{propuni})
valid in the unitary gauge. 
Eq.~(\ref{proprxi}) however, when compared to eq.~(\ref{propuni}), 
has the advantage of being applicable both to massive and to massless 
vector bosons.
Furthermore, in the limit $q \rightarrow \infty$, 
it goes to zero, thus the UV behaviour of the theory
is not spoiled by the choice of this gauge.
Actually, we worked both in the generalized 
$R_\xi$ gauges and in the unitary gauge,
and we explicitly verified that the ${\rm R_2}$ effective vertices 
obtained in the latter have
expressions far more complicated~\cite{ultimo}, 
as expected on the basis of the previous 
considerations.\footnote{When working in the Unitary gauge, at
the purpose of calculating the ${\rm R_2}$ (and the ${\rm R_1}$)
effective vertices,
we took the limit  $\xi, \, \xi_{Z} \rightarrow \infty$ and
$\xi_A \rightarrow 1$  before the integration, according to the 
observations in~\cite{ultimo}.}

\section{Procedure followed in the calculation of the ${\rm R_2}$ effective vertices and structure of the code}
\label{program}

We worked in the SM of the 
EW interactions, and we calculated in an analytical way, 
by using FORM~\cite{form}, the general ${\rm R_2}$ contributions 
corresponding to all possible 1-particle 
irreducible graphs, with up to 4 external legs, written in
terms of $generic$ scalars $s$, vectors $v$ and fermions $f$
(as explained in the Introduction, we denote the graphs where 
 generic fields, like $s$, $v$ and $f$, appear instead of specific 
 particles, like $H$, $Z$, $e^-$, etc., as 
 $generic$ $diagrams$ throughout this paper).  

The starting point for the evaluation of the ${\rm R_2}$ contribution 
to each generic diagram is the corresponding ${\rm R_2}$ integrand expression
(under the integral sign in eq.~(\ref{eqr2})),
built as suggested in Section~\ref{section2}, once fixed
a gauge, on the basis of the considerations in Section~\ref{section3}. 
Feynman rules written in terms of generic $s$, 
$v$ and $f$ according to Denner conventions~\cite{denner}, enter
the numerator of this integrand. 
The integration  
can then be performed analitically by using conventional techniques,
involving Feynman parametrization followed by Wick rotation. 
The basic formulas for the angular and radial integrations are given
by~\cite{uni0}
\begin{eqnarray}
\int d \Omega_d & = & \frac{2 \pi^{d/2}}{\Gamma(d/2)} \, , \\
\int_0^{\infty} dq \frac{q^\beta}{(q^2 + X)^\alpha} &=& \frac{1}{2} 
\frac{\Gamma\left(\frac{\beta +1}{2}\right) \Gamma\left(\alpha - \frac{\beta+1}{2}\right)}{
\Gamma(\alpha) X^{\alpha - \frac{\beta+1}{2}}} \, .
\end{eqnarray} 
The integration always leads to the disappearence of all
divergencies involving negative powers of $\epsilon$. 
One can thus safely take the limit $\epsilon \rightarrow 0$
just after it.

By following this procedure, we obtain the $\rm R_2$ analytical
expressions corresponding
to the generic diagrams
listed in the 
$\texttt{xxxxgentop.h}$ 
files present in the package, 
with \texttt{xxxx}~=~\texttt{ss}, \texttt{vs}, \texttt{vv}, \texttt{ff},
\texttt{sff}, \texttt{vff},
\texttt{sss}, 
\texttt{vss}, \texttt{svv}, \texttt{vvv}, \texttt{ssss}, \texttt{ssvv}
and \texttt{vvvv}. 
The name of each of these files refers to the
nature of the external particles and the results are presented 
generic diagram by generic diagram (e.g. the file 
$\texttt{vssgentop.h}$ includes all 
${\rm R_2}$ contributions 
to the 3-point functions including
one vector and two scalars, generic diagram by generic diagram, 
where each generic diagram differs from the others according to the 
topology and/or to the nature
of the fields internal to the loop).

Only the generic diagrams giving rise to an ${\rm R_2}$ contribution different
from zero are listed. 
The main criterion for establishing if a diagram can contribute or not to
${\rm R_2}$ is power counting in the loop momentum. 
To every integral of the type
\begin{eqnarray}
\int d^d \bar{q} \, {\tilde{q}}^{2l} \, \frac{q_{\mu_1}...q_{\mu_{2s}}}{\bar{D}_0
..\bar{D}_m}
\end{eqnarray}
we associated an index $d^{\,\prime} = l+s+1-m$. Integrals with $d^{\,\prime} \ge 0$ can give a
contribution, while integrals with $d^{\,\prime} < 0$ vanish~\cite{aguila}. 
By checking in each
diagram the number of available loop momenta in the numerator and by
counting the number of denominators, we could say 
beforehand if they may contribute 
or not. 
All gauges in the $R_\xi$ class, included the 't Hooft Feynman one,
have the same power counting properties.
On the other hand, in the unitary gauge 
the situation is different 
and
more diagrams can contribute than in the previous case, 
at least as for power counting, 
since the piece of the propagator~(\ref{propuni}) proportional to 
$q^\mu q^\nu$ just includes two more powers of 
momentum in the numerator and thus passes more easily 
the power counting test.
This is counterbalanced by the fact that, due to the reduced 
number of degrees of freedom, many diagrams are absent (e.g. 
all those including a {\texttt{ssv}} vertex, since two physical
$s$ don't couple with a $v$ and diagrams including unphysical
Goldstone $s$ are absent).
Another criterion is the existence or not  of terms
proportional to $\epsilon$ or ${\tilde q}^2$ in $\bar{N}(\bar{q})$. 
In the case of ghosts for example, the loop momenta in the 
ghost-ghost-vector Feynman rules have an
index that comes from an external vector and therefore is 4 
dimensional. When such an object is contracted with a loop
momentum, it cannot produce  ${\tilde q}^2$ terms and 
thus can not contribute to $\rm R_2$ terms.\footnote{
Ghost loops do not contribute to
$\rm R_2$ terms. These loops however may contribute to the $\rm R_1$ part 
of the amplitude, where have to be taken properly into account.}

\begin{figure}[h!]
\begin{center}
  \begin{picture}(550,250) 
    \SetScale{0.4}
    \SetWidth{0.5}
    \SetColor{Black}


    \SetOffset(-60,240)
    \DashLine(65,0)(133,0){8}
    \DashLine(65,-68)(133,-68){8}
    \DashLine(201,0)(269,0){8}
    \DashLine(201,-68)(269,-68){8}
    \DashLine (133,0)(201,0){8}
    \DashLine (133,-68)(201,-68){8}
    \Photon (201,0)(201,-68){3.5}{5}
    \Photon (133,0)(133,-68){3.5}{5}

    \Vertex(133,0){4}
    \Vertex(201,0){4}
    \Vertex(133,-68){4}
    \Vertex(201,-68){4}

    \SetOffset(50,240)
    \DashLine(65,0)(133,0){8}
    \DashLine(65,-68)(133,-68){8}
    \DashLine(201,0)(269,0){8}
    \DashLine(201,-68)(269,-68){8}
    \ArrowLine (133,0)(201,0)
    \ArrowLine (201,-68)(133,-68)
    \ArrowLine (201,0)(201,-68)
    \ArrowLine (133,-68)(133,0)

    \Vertex(133,0){4}
    \Vertex(201,0){4}
    \Vertex(133,-68){4}
    \Vertex(201,-68){4}

    \SetOffset(160,240)
    \DashLine(65,0)(133,0){8}
    \DashLine(65,-68)(133,-68){8}
    \DashLine(201,-34)(269,0){8}
    \DashLine(201,-34)(269,-68){8}
    \DashLine (133,0)(201,-34){8}
    \DashLine (133,-68)(201,-34){8}
    \Photon (133,0)(133,-68){3.5}{5}

    \Vertex(133,0){4}
    \Vertex(201,-34){4}
    \Vertex(133,-68){4}

    \SetOffset(270,240)
    \DashLine(65,0)(133,0){8}
    \DashLine(65,-68)(133,-68){8}
    \DashLine(201,-34)(269,0){8}
    \DashLine(201,-34)(269,-68){8}
    \Photon (133,0)(201,-34){3.5}{5}
    \Photon (133,-68)(201,-34){3.5}{5}
    \DashLine (133,0)(133,-68){8}

    \Vertex(133,0){4}
    \Vertex(201,-34){4}
    \Vertex(133,-68){4}

    \SetOffset(380,240)
    \DashLine(65,0)(133,-34){8}
    \DashLine(65,-68)(133,-34){8}
    \DashLine(201,-34)(269,0){8}
    \DashLine(201,-34)(269,-68){8}
    \PhotonArc(167,-34)(34.0,0,180){3.5}{8}
    \PhotonArc(167,-34)(34.0,180,360){3.5}{8}

    \Vertex(133,-34){4}
    \Vertex(201,-34){4}


    \SetOffset(-60,160)
    \DashLine(65,0)(133,0){8}
    \DashLine(65,-68)(133,-68){8}
    \Photon(201,0)(269,0){3.5}{5}
    \Photon(201,-68)(269,-68){3.5}{5}
    \DashLine (133,0)(201,0){8}
    \DashLine (133,-68)(201,-68){8}
    \DashLine (201,0)(201,-68){8}
    \Photon (133,0)(133,-68){3.5}{5}

    \Vertex(133,0){4}
    \Vertex(201,0){4}
    \Vertex(133,-68){4}
    \Vertex(201,-68){4}

    \SetOffset(50,160)
    \DashLine(65,0)(133,0){8}
    \DashLine(65,-68)(133,-68){8}
    \Photon(201,0)(269,0){3.5}{5}
    \Photon(201,-68)(269,-68){3.5}{5}
    \Photon (133,0)(201,0){3.5}{5}
    \Photon (133,-68)(201,-68){3.5}{5}
    \Photon (201,0)(201,-68){3.5}{5}
    \DashLine (133,0)(133,-68){8}

    \Vertex(133,0){4}
    \Vertex(201,0){4}
    \Vertex(133,-68){4}
    \Vertex(201,-68){4}

    \SetOffset(160,160)
    \DashLine(65,0)(133,0){8}
    \DashLine(65,-68)(133,-68){8}
    \Photon(201,0)(269,0){3.5}{5}
    \Photon(201,-68)(269,-68){3.5}{5}
    \ArrowLine (133,0)(201,0)
    \ArrowLine (201,-68)(133,-68)
    \ArrowLine (201,0)(201,-68)
    \ArrowLine (133,-68)(133,0)

    \Vertex(133,0){4}
    \Vertex(201,0){4}
    \Vertex(133,-68){4}
    \Vertex(201,-68){4}

    \SetOffset(270,160)
    \Photon(65,0)(133,0){3.5}{5}
    \DashLine(65,-68)(133,-68){8}
    \DashLine(201,0)(269,0){8}
    \Photon(201,-68)(269,-68){3.5}{5}
    \ArrowLine (133,0)(201,0)
    \ArrowLine (201,-68)(133,-68)
    \ArrowLine (201,0)(201,-68)
    \ArrowLine (133,-68)(133,0)

    \Vertex(133,0){4}
    \Vertex(201,0){4}
    \Vertex(133,-68){4}
    \Vertex(201,-68){4}

    \SetOffset(380,160)
    \DashLine(65,0)(133,-34){8}
    \DashLine(65,-68)(133,-34){8}
    \Photon(201,-34)(269,0){3.5}{5}
    \Photon(201,-34)(269,-68){3.5}{5}
    \PhotonArc(167,-34)(34.0,0,180){3.5}{8}
    \PhotonArc(167,-34)(34.0,180,360){3.5}{8}

    \Vertex(133,-34){4}
    \Vertex(201,-34){4}

    \SetOffset(-60,100)
    \DashLine(65,0)(133,0){8}
    \DashLine(65,-68)(133,-68){8}
    \Photon(201,-34)(269,0){3.5}{5}
    \Photon(201,-34)(269,-68){3.5}{5}
    \DashLine (133,0)(201,-34){8}
    \DashLine (133,-68)(201,-34){8}
    \Photon (133,0)(133,-68){3.5}{5}

    \Vertex(133,0){4}
    \Vertex(201,-34){4}
    \Vertex(133,-68){4}

    \SetOffset(50,100)
    \DashLine(65,0)(133,0){8}
    \DashLine(65,-68)(133,-68){8}
    \Photon(201,-34)(269,0){3.5}{5}
    \Photon(201,-34)(269,-68){3.5}{5}
    \Photon (133,0)(201,-34){3.5}{5}
    \Photon (133,-68)(201,-34){3.5}{5}
    \DashLine (133,0)(133,-68){8}

    \Vertex(133,0){4}
    \Vertex(201,-34){4}
    \Vertex(133,-68){4}

    \SetOffset(160,100)
    \Photon(65,0)(133,0){3.5}{5}
    \DashLine(65,-68)(133,-68){8}
    \Photon(201,-34)(269,0){3.5}{5}
    \DashLine(201,-34)(269,-68){8}
    \DashLine (133,0)(201,-34){8}
    \Photon (133,-68)(201,-34){3.5}{5}
    \DashLine (133,0)(133,-68){8}

    \Vertex(133,0){4}
    \Vertex(201,-34){4}
    \Vertex(133,-68){4}

    \SetOffset(270,100)
    \Photon(65,0)(133,0){3.5}{5}
    \DashLine(65,-68)(133,-68){8}
    \Photon(201,-34)(269,0){3.5}{5}
    \DashLine(201,-34)(269,-68){8}
    \Photon (133,0)(201,-34){3.5}{5}
    \DashLine (133,-68)(201,-34){8}
    \Photon (133,0)(133,-68){3.5}{5}

    \Vertex(133,0){4}
    \Vertex(201,-34){4}
    \Vertex(133,-68){4}

    \SetOffset(380,100)
    \Photon(65,0)(133,0){3.5}{5}
    \Photon(65,-68)(133,-68){3.5}{5}
    \DashLine(201,-34)(269,0){8}
    \DashLine(201,-34)(269,-68){8}
    \Photon (133,0)(201,-34){3.5}{5}
    \Photon (133,-68)(201,-34){3.5}{5}
    \Photon (133,0)(133,-68){3.5}{5}

    \Vertex(133,0){4}
    \Vertex(201,-34){4}
    \Vertex(133,-68){4}


    \SetOffset(0,20)
    \Photon(65,0)(133,0){3.5}{5}
    \Photon(65,-68)(133,-68){3.5}{5}
    \Photon(201,0)(269,0){3.5}{5}
    \Photon(201,-68)(269,-68){3.5}{5}
    \Photon (133,0)(201,0){3.5}{5}
    \Photon (133,-68)(201,-68){3.5}{5}
    \Photon (201,0)(201,-68){3.5}{5}
    \Photon (133,0)(133,-68){3.5}{5}

    \Vertex(133,0){4}
    \Vertex(201,0){4}
    \Vertex(133,-68){4}
    \Vertex(201,-68){4}

    \SetOffset(110,20)
    \Photon(65,0)(133,0){3.5}{5}
    \Photon(65,-68)(133,-68){3.5}{5}
    \Photon(201,0)(269,0){3.5}{5}
    \Photon(201,-68)(269,-68){3.5}{5}
    \ArrowLine (133,0)(201,0)
    \ArrowLine (201,-68)(133,-68)
    \ArrowLine (201,0)(201,-68)
    \ArrowLine (133,-68)(133,0)

    \Vertex(133,0){4}
    \Vertex(201,0){4}
    \Vertex(133,-68){4}
    \Vertex(201,-68){4}

    \SetOffset(220,20)
    \Photon(65,0)(133,0){3.5}{5}
    \Photon(65,-68)(133,-68){3.5}{5}
    \Photon(201,-34)(269,0){3.5}{5}
    \Photon(201,-34)(269,-68){3.5}{5}
    \Photon (133,0)(201,-34){3.5}{5}
    \Photon (133,-68)(201,-34){3.5}{5}
    \Photon (133,0)(133,-68){3.5}{5}

    \Vertex(133,0){4}
    \Vertex(201,-34){4}
    \Vertex(133,-68){4}

    \SetOffset(330,20)
    \Photon(65,0)(133,-34){3.5}{5}
    \Photon(65,-68)(133,-34){3.5}{5}
    \Photon(201,-34)(269,0){3.5}{5}
    \Photon(201,-34)(269,-68){3.5}{5}
    \PhotonArc(167,-34)(34.0,0,180){3.5}{8}
    \PhotonArc(167,-34)(34.0,180,360){3.5}{8}

    \Vertex(133,-34){4}
    \Vertex(201,-34){4}







  \end{picture}
\end{center}
\caption{\label{figura5} Non null contributions 
to the {\texttt{ssss}}, {\texttt{ssvv}} 
and {\texttt{vvvv}} 
${\mathrm R_2}$ effective 
vertices in the generalized $R_\xi$ gauges, with generic finite
$\xi$, $\xi_Z$, $\xi_A$. 
The corresponding analytical formulas associated to selected
topologies of
each generic
diagram are included in the files $\texttt{xxxxgentop.h}$, 
with $\texttt{xxxx}$ = $\texttt{ssss}$, $\texttt{ssvv}$ and 
$\texttt{vvvv}$,
respectively. 
For all diagrams including at least a $\texttt{vvvv}$ vertex 
see the comments in Fig.~\protect\ref{figura2}.
}
\end{figure}
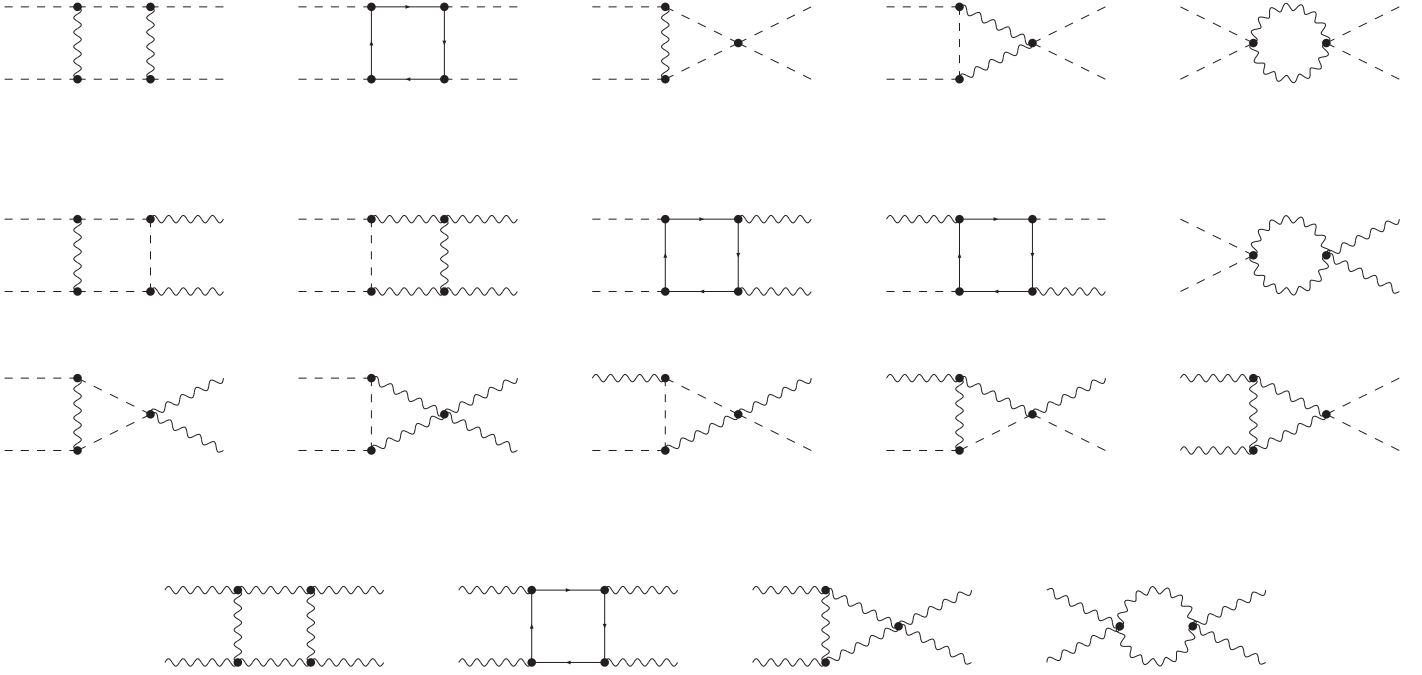

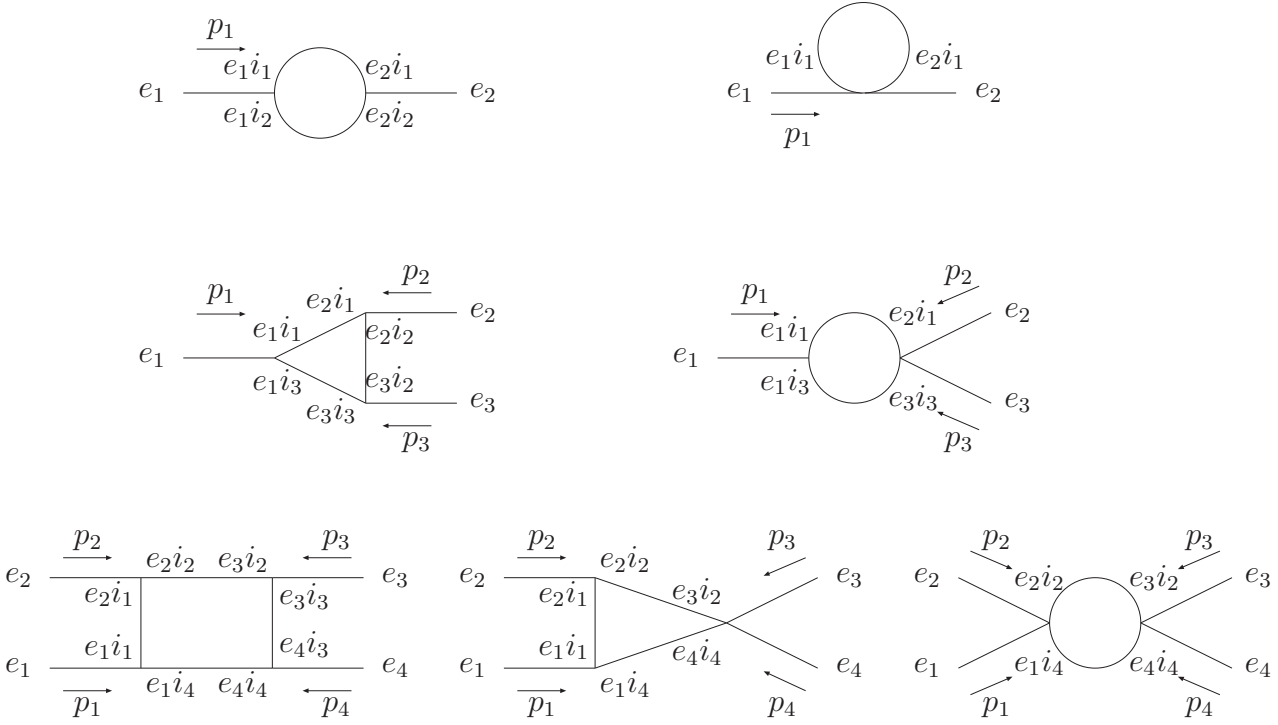
\begin{figure}[h!]
\begin{center}
  \begin{picture}(550,250) 
    \SetOffset(0,242)
    \SetScale{0.5}
    \SetWidth{0.5}
    \SetColor{Black}
    \Line(65,-34)(133,-34)
    \BCirc(167,-34){34.0}

    \Line(201,-34)(269,-34)
    \Text(47,7)[]{{\Black{$p_1$}}}
    \LongArrow(75,-1)(110,-1)  
    \Text(21,-17)[]{$e_1$}
    \Text(145,-17)[]{$e_2$}
    \Text(110,-7)[]{$e_2i_1$}
    \Text(110,-25)[]{$e_2i_2$}
    \Text(57,-25)[]{$e_1i_2$}
    \Text(57,-7)[]{$e_1i_1$}

    \SetOffset(220,242)

    \Line(65,-34)(133,-34)
    \BCirc(134,0){34.0}
    \Line(135,-34)(203,-34)
    \Text(40,-3)[]{{\Black{$e_1i_1$}}}
    \Text(96,-3)[]{{\Black{$e_2i_1$}}}
    \LongArrow(65,-50)(100,-50)  
    \Text(43,-34)[]{{\Black{$p_1$}}}
    \Text(21,-17)[]{$e_1$}
    \Text(114,-17)[]{$e_2$}

    \SetOffset(0,142)
    \Line(65,-34)(133,-34)
    \Line (133,-34)(201,0)
    \Line (133,-34)(201,-68)
    \Line (201,0)(201,-68)
    \Line(201,0)(269,0)
    \Line(201,-68)(269,-68)
    \Text(47,7)[]{{\Black{$p_1$}}}
    \LongArrow(75,-1)(110,-1)  
    \Text(21,-17)[]{$e_1$}
    \Text(145,0)[]{$e_2$}
    \Text(145,-34)[]{$e_3$}

    \LongArrow(250,15)(215,15)  
    \Text(120,14)[]{{\Black{$p_2$}}}

    \LongArrow(250,-85)(215,-85)  
    \Text(120,-49)[]{{\Black{$p_3$}}}

    \Text(110,-7)[]{$e_2i_2$}
    \Text(110,-25)[]{$e_3i_2$}
    \Text(68,-26)[]{$e_1i_3$}
    \Text(68,-6)[]{$e_1i_1$}
    \Text(88,5)[]{$e_2i_1$}
    \Text(88,-37)[]{$e_3i_3$}

    \SetOffset(200,142)
    \Line(65,-34)(133,-34)
    \BCirc(167,-34){34.0}


    \Line(201,-34)(269,0)
    \Line(201,-34)(269,-68)
    \Text(47,7)[]{{\Black{$p_1$}}}
    \LongArrow(75,-1)(110,-1)  
    \Text(21,-17)[]{$e_1$}
    \Text(145,0)[]{$e_2$}
    \Text(145,-34)[]{$e_3$}

    \LongArrow(260,20)(230,7)  
    \Text(123,14)[]{{\Black{$p_2$}}}

    \LongArrow(260,-88)(230,-75)  
    \Text(123,-49)[]{{\Black{$p_3$}}}

    \Text(58,-26)[]{$e_1i_3$}
    \Text(58,-6)[]{$e_1i_1$}
    \Text(106,0)[]{$e_2i_1$}
    \Text(106,-32)[]{$e_3i_3$}

    \SetOffset(-50,42)
    \Line(65,0)(133,0)
    \Line(65,-68)(133,-68)
    \Line(231,0)(299,0)
    \Line(231,-68)(299,-68)
    \Line (133,0)(231,0)
    \Line (133,-68)(231,-68)
    \Line (231,0)(231,-68)
    \Line (133,0)(133,-68)

    \Text(21,-34)[]{$e_1$}
    \Text(21,0)[]{$e_2$}
    \Text(162,0)[]{$e_3$}
    \Text(162,-34)[]{$e_4$}

    \LongArrow(75,-85)(110,-85)  
    \Text(47,-49)[]{{\Black{$p_1$}}}

    \LongArrow(75,15)(110,15)  
    \Text(47,14)[]{{\Black{$p_2$}}}

    \LongArrow(290,15)(255,15)  
    \Text(140,14)[]{{\Black{$p_3$}}}

    \LongArrow(290,-85)(255,-85)  
    \Text(140,-49)[]{{\Black{$p_4$}}}

    \Text(128,-7)[]{$e_3i_3$}
    \Text(128,-25)[]{$e_4i_3$}
    \Text(55,-26)[]{$e_1i_1$}
    \Text(55,-6)[]{$e_2i_1$}
    \Text(105,7)[]{$e_3i_2$}
    \Text(78,7)[]{$e_2i_2$}
    \Text(105,-40)[]{$e_4i_4$}
    \Text(78,-40)[]{$e_1i_4$}

    \SetOffset(120,42)
    \Line(65,0)(133,0)
    \Line(65,-68)(133,-68)
    \Line(231,-34)(299,0)
    \Line(231,-34)(299,-68)
    \Line (133,0)(231,-34)
    \Line (133,-68)(231,-34)
    \Line (133,0)(133,-68)

    \Text(21,-34)[]{$e_1$}
    \Text(21,0)[]{$e_2$}
    \Text(162,0)[]{$e_3$}
    \Text(162,-34)[]{$e_4$}

    \LongArrow(75,-85)(110,-85)  
    \Text(47,-49)[]{{\Black{$p_1$}}}

    \LongArrow(75,15)(110,15)  
    \Text(47,14)[]{{\Black{$p_2$}}}

    \LongArrow(290,15)(260,2)  
    \Text(137,14)[]{{\Black{$p_3$}}}

    \LongArrow(290,-85)(260,-71)  
    \Text(137,-49)[]{{\Black{$p_4$}}}

    \Text(105,-5)[]{$e_3i_2$}
    \Text(105,-28)[]{$e_4i_4$}
    \Text(55,-26)[]{$e_1i_1$}
    \Text(55,-6)[]{$e_2i_1$}
    \Text(78,7)[]{$e_2i_2$}
    \Text(78,-40)[]{$e_1i_4$}

    \SetOffset(290,42)
    \Line(65,0)(133,-34)
    \Line(65,-68)(133,-34)
    \BCirc(167,-34){34.0}


    \Line(201,-34)(269,0)
    \Line(201,-34)(269,-68)

    \LongArrow(74,20)(104,7)  
    \Text(47,14)[]{{\Black{$p_2$}}}

    \LongArrow(74,-88)(104,-75)  
    \Text(47,-49)[]{{\Black{$p_1$}}}

    \Text(21,0)[]{$e_2$}
    \Text(21,-34)[]{$e_1$}
    \Text(145,0)[]{$e_3$}
    \Text(145,-34)[]{$e_4$}

    \LongArrow(260,20)(230,7)  
    \Text(123,14)[]{{\Black{$p_3$}}}

    \LongArrow(260,-88)(230,-75)  
    \Text(123,-49)[]{{\Black{$p_4$}}}

    \Text(63,-32)[]{$e_1i_4$}
    \Text(63,0)[]{$e_2i_2$}
    \Text(106,0)[]{$e_3i_2$}
    \Text(106,-32)[]{$e_4i_4$}

  \end{picture}
\end{center}
\caption{\label{figura1} 2-,3- and 4-point 
1-particle irreducible topologies {\it explicitly} considered
for the calculation of the $\rm R_2$ contributions included in the  
{\texttt{xxxxgentop.h}} files. The symbols $e_j$ (\texttt{ext}$j$\texttt{fla}
in our files), with $j$ = 1,2,3,4, denote
external particles. Two indices are associated to each internal particle
$i_k$ (k = 1,2,3,4), corresponding to the two vertices to which it is 
connected. So $e_j i_k$ (\texttt{e}$j$\texttt{int}$k$\texttt{fla}
in our files) denotes the $k$-th internal particle {\it incoming} 
in the vertex in which also the $j$-th external particle is entering.
According to the notation used to write down the Feynman rules,
all momenta in all vertices are supposed to be incoming.
Different topologies have been obtained from these ones by non-cyclic 
permutations of the external particles, together with their momenta and 
their Lorentz indices (omitted for simplicity in this figure), if present, 
for each fixed configuration of the internal ones.}
\end{figure}

The generic diagrams giving a non-null $\rm R_2$ contribution in the
$R_\xi$ gauges  
are shown in Fig.~\ref{figura3},~\ref{figura4}~and~\ref{figura5},
where each of them is presented in a single topological 
configuration, just for compactness.
In fact, only selected topologies have been explicitly 
dressed to generate generic diagrams, i.e.
we explicitly considered 
only one
topology associated to
each fixed number of internal and external legs, 
and then different topologies (and the corresponding generic diagrams) 
have been obtained from the first
ones by proper non-cyclic permutations of the momenta 
of the external particles, at a fixed configuration of the internal particles. 
In case of external bosons, besides the momenta,
even the Lorentz indices have to be properly permuted,
to obtain one topology/diagram from another.  
The schemes showing our convention corresponding to the topologies
explicitly considered for 2-leg (bubbles and tadpoles), 
3-leg (triangles and bubbles) 
and 4-leg (boxes, triangles and bubbles) diagrams, 
are shown in Fig.~\ref{figura1}.

Each $\mathrm R_2$ contribution included 
in a $\texttt{xxxxgentop.h}$ file
is labelled by $\texttt{EFFVERyyyyLzzzz}$, where $\texttt{yyyy}$ correspond 
to the {\it ordered} list of the generic 
external particles (i.e. the first particle corresponds to the $p_1$ momentum, 
the second one to $p_2$, etc.....), $\texttt{L}$ stays for ``loop'', 
and $\texttt{zzzz}$ is the {\it ordered} list of the generic 
particles running inside the loop, where
the first internal particle is, by convention, 
the one joining external particle 1 with external particle 2, and so on,
up to the last internal particle joining the last external particle to
the first external one.    

Furthermore, the $\mathrm R_2$ contribution corresponding to each 
generic diagram is presented as an explicit FORM function of the external 
and internal particles. As shown in Fig.~\ref{figura1}, 
we associated two labels to every internal particle, each of them 
in correspondence with the different vertex to which it is joined. 
The reason of this choice is better accomodating charged particles and 
fermions (an internal fermionic line corresponds to both an antifermion
emerging from a vertex and to a fermion entering in another vertex), by 
taking into account that in each vertex all momenta are supposed 
to be incoming. This notation seems a little bit redundant, however
we found it very useful in making clearer the code.

\begin{figure}[h!]
\begin{center}
  \begin{picture}(550,270) 
    \SetScale{0.4}
    \SetWidth{0.5}
    \SetColor{Black}


    \SetOffset(0,260)
    \DashLine(65,-34)(133,-34){8}
    \PhotonArc(167,-34)(34.0,0,180){3.5}{8}
    \DashCArc(167,-34)(34.0,180,360){8}
    \Vertex(133,-34){4}
    \Vertex(201,-34){4}
    \DashLine(201,-34)(269,-34){8}

    \SetOffset(110,260)
    \DashLine(65,-34)(133,-34){8}
    \PhotonArc(167,-34)(34.0,0,180){3.5}{8}
    \PhotonArc(167,-34)(34.0,180,360){3.5}{8}
    \Vertex(133,-34){4}
    \Vertex(201,-34){4}
    \DashLine(201,-34)(269,-34){8}




    \SetOffset(220,260)
    \DashLine(65,-34)(133,-34){8}
    \ArrowArc(167,-34)(34.0,0,180)
    \ArrowArc(167,-34)(34.0,180,360)
    \Vertex(133,-34){4}
    \Vertex(201,-34){4}
    \DashLine(201,-34)(269,-34){8}

    \SetOffset(330,260)
    \DashLine(65,-34)(133,-34){8}
    \PhotonArc(134,4)(34.0,0,360){3.5}{15}
    \DashLine(135,-34)(203,-34){8}

    \Vertex(134,-34){4}


    \SetOffset(110,180)
    \Photon(65,-34)(133,-34){3.5}{5}
    \PhotonArc(167,-34)(34.0,0,180){3.5}{8}
    \DashCArc(167,-34)(34.0,180,360){8}
    \Vertex(133,-34){4}
    \Vertex(201,-34){4}
    \DashLine(201,-34)(269,-34){8}

    \SetOffset(220,180)
    \Photon(65,-34)(133,-34){3.5}{5}
    \PhotonArc(167,-34)(34.0,0,180){3.5}{8}
    \PhotonArc(167,-34)(34.0,180,360){3.5}{8}
    \Vertex(133,-34){4}
    \Vertex(201,-34){4}
    \DashLine(201,-34)(269,-34){8}


    \SetOffset(50,100)
    \Photon(65,-34)(133,-34){3.5}{5}
    \PhotonArc(167,-34)(34.0,0,180){3.5}{8}
    \PhotonArc(167,-34)(34.0,180,360){3.5}{8}
    \Vertex(133,-34){4}
    \Vertex(201,-34){4}
    \Photon(201,-34)(269,-34){3.5}{5}

    \SetOffset(160,100)
    \Photon(65,-34)(133,-34){3.5}{5}
    \ArrowArc(167,-34)(34.0,0,180)
    \ArrowArc(167,-34)(34.0,180,360)
    \Vertex(133,-34){4}
    \Vertex(201,-34){4}
    \Photon(201,-34)(269,-34){3.5}{5}

    \SetOffset(270,100)
    \Photon(65,-34)(133,-34){3.5}{5}
    \PhotonArc(134,4)(34.0,0,360){3.5}{15}
    \Photon(135,-34)(203,-34){3.5}{5}
    \Vertex(135,-34){4}


    \SetOffset(160,20)
    \ArrowLine(65,-34)(133,-34)
    \ArrowArc(167,-34)(34.0,180,360)
    \PhotonArc(167,-34)(34.0,0,180){3.5}{8}
    \Vertex(133,-34){4}
    \Vertex(201,-34){4}
    \ArrowLine(201,-34)(269,-34)

  \end{picture}
\end{center}
\caption{\label{figura3} Non null contributions 
to the {\texttt{ss}}, {\texttt{vs}}, {\texttt{vv}} and {\texttt{ff}} 
${\mathrm R_2}$ effective vertices 
in the generalized $R_\xi$ gauges, with generic finite $\xi$, $\xi_Z$, $\xi_A$. 
The corresponding analytical formulas associated to selected
topologies of
each generic
diagram are included in the files $\texttt{xxxxgentop.h}$, 
with $\texttt{xxxx}$ = $\texttt{ss}$, $\texttt{vs}$, $\texttt{vv}$ 
and $\texttt{ff}$, 
respectively.}
\end{figure}
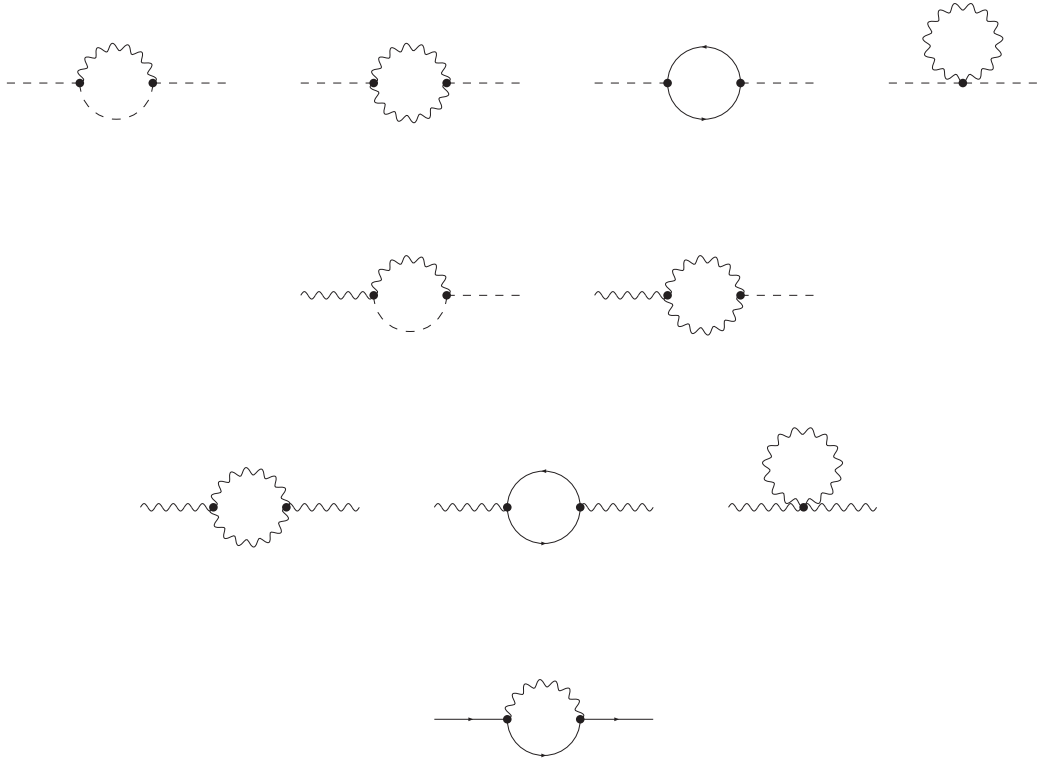

\begin{figure}[h!]
\begin{center}
  \begin{picture}(550,430) 
    \SetScale{0.4}
    \SetWidth{0.5}
    \SetColor{Black}


    \SetOffset(0,420)
    \DashLine(65,-34)(133,-34){8}
    \ArrowLine(201,0)(269,0)
    \ArrowLine(269,-68)(201,-68)
    \DashLine (133,-34)(201,0){8}
    \ArrowLine (201,-68)(201,0)
    \Photon (133,-34)(201,-68){3.5}{5}
    \Vertex(133,-34){4}
    \Vertex(201,0){4}
    \Vertex(201,-68){4}

    \SetOffset(110,420)
    \DashLine(65,-34)(133,-34){8}
    \ArrowLine(201,0)(269,0)
    \ArrowLine(269,-68)(201,-68)
    \Photon (133,-34)(201,0){3.5}{5}
    \ArrowLine (201,-68)(201,0)
    \DashLine (133,-34)(201,-68){8}
    \Vertex(133,-34){4}
    \Vertex(201,0){4}
    \Vertex(201,-68){4}

    \SetOffset(220,420)
    \DashLine(65,-34)(133,-34){8}
    \ArrowLine(201,0)(269,0)
    \ArrowLine(269,-68)(201,-68)
    \ArrowLine (133,-34)(201,0)
    \DashLine (201,0)(201,-68){8}
    \ArrowLine (201,-68)(133,-34)
    \Vertex(133,-34){4}
    \Vertex(201,0){4}
    \Vertex(201,-68){4}

    \SetOffset(330,420)
    \DashLine(65,-34)(133,-34){8}
    \ArrowLine(201,0)(269,0)
    \ArrowLine(269,-68)(201,-68)
    \ArrowLine (133,-34)(201,0)
    \Photon (201,0)(201,-68){3.5}{5}
    \ArrowLine (201,-68)(133,-34)
    \Vertex(133,-34){4}
    \Vertex(201,0){4}
    \Vertex(201,-68){4}


    \SetOffset(50,340)
    \Photon(65,-34)(133,-34){3.5}{5}
    \ArrowLine(201,0)(269,0)
    \ArrowLine(269,-68)(201,-68)
    \Photon (133,-34)(201,0){3.5}{5}
    \ArrowLine (201,-68)(201,0)
    \Photon (133,-34)(201,-68){3.5}{5}
    \Vertex(133,-34){4}
    \Vertex(201,0){4}
    \Vertex(201,-68){4}

    \SetOffset(160,340)
    \Photon(65,-34)(133,-34){3.5}{5}
    \ArrowLine(201,0)(269,0)
    \ArrowLine(269,-68)(201,-68)
    \ArrowLine (133,-34)(201,0)
    \DashLine (201,0)(201,-68){8}
    \ArrowLine (201,-68)(133,-34)
    \Vertex(133,-34){4}
    \Vertex(201,0){4}
    \Vertex(201,-68){4}

    \SetOffset(270,340)
    \Photon(65,-34)(133,-34){3.5}{5}
    \ArrowLine(201,0)(269,0)
    \ArrowLine(269,-68)(201,-68)
    \ArrowLine (133,-34)(201,0)
    \Photon (201,0)(201,-68){3.5}{5}
    \ArrowLine (201,-68)(133,-34)
    \Vertex(133,-34){4}
    \Vertex(201,0){4}
    \Vertex(201,-68){4}


    \SetOffset(0,260)
    \DashLine(65,-34)(133,-34){8}
    \DashLine(201,0)(269,0){8}
    \DashLine(201,-68)(269,-68){8}
    \DashLine (133,-34)(201,0){8}
    \DashLine (201,0)(201,-68){8}
    \Photon (133,-34)(201,-68){3.5}{5}
    \Vertex(133,-34){4}
    \Vertex(201,0){4}
    \Vertex(201,-68){4}

    \SetOffset(110,260)
    \DashLine(65,-34)(133,-34){8}
    \DashLine(201,0)(269,0){8}
    \DashLine(201,-68)(269,-68){8}
    \DashLine (133,-34)(201,0){8}
    \Photon (201,0)(201,-68){3.5}{5}
    \Photon (133,-34)(201,-68){3.5}{5}
    \Vertex(133,-34){4}
    \Vertex(201,0){4}
    \Vertex(201,-68){4}

    \SetOffset(220,260)
    \DashLine(65,-34)(133,-34){8}
    \DashLine(201,0)(269,0){8}
    \DashLine(201,-68)(269,-68){8}
    \ArrowLine (133,-34)(201,0)
    \ArrowLine (201,0)(201,-68)
    \ArrowLine (201,-68)(133,-34)
    \Vertex(133,-34){4}
    \Vertex(201,0){4}
    \Vertex(201,-68){4}

    \SetOffset(330,260)
    \PhotonArc(167,-34)(34.0,0,180){3.5}{8}
    \PhotonArc(167,-34)(34.0,180,360){3.5}{8}
    \DashLine(65,-34)(133,-34){8}
    \DashLine(201,-34)(269,0){8}
    \DashLine(201,-34)(269,-68){8}
    \Vertex(133,-34){4}
    \Vertex(201,-34){4}


    \SetOffset(-60,180)
    \Photon(65,-34)(133,-34){3.5}{5}
    \DashLine(201,0)(269,0){8}
    \DashLine(201,-68)(269,-68){8}
    \DashLine(133,-34)(201,0){8}
    \Photon(201,0)(201,-68){3.5}{5}
    \DashLine(133,-34)(201,-68){8}
    \Vertex(133,-34){4}
    \Vertex(201,0){4}
    \Vertex(201,-68){4}

    \SetOffset(50,180)
    \Photon(65,-34)(133,-34){3.5}{5}
    \DashLine(201,0)(269,0){8}
    \DashLine(201,-68)(269,-68){8}
    \Photon (133,-34)(201,0){3.5}{5}
    \DashLine (201,0)(201,-68){8}
    \Photon (133,-34)(201,-68){3.5}{5}
    \Vertex(133,-34){4}
    \Vertex(201,0){4}
    \Vertex(201,-68){4}

    \SetOffset(160,180)
    \Photon(65,-34)(133,-34){3.5}{5}
    \DashLine(201,0)(269,0){8}
    \DashLine(201,-68)(269,-68){8}
    \ArrowLine (133,-34)(201,0)
    \ArrowLine (201,0)(201,-68)
    \ArrowLine (201,-68)(133,-34)
    \Vertex(133,-34){4}
    \Vertex(201,0){4}
    \Vertex(201,-68){4}

    \SetOffset(270,180)
    \DashCArc(167,-34)(34.0,0,180){8}
    \PhotonArc(167,-34)(34.0,180,360){3.5}{8}
    \DashLine(65,-34)(133,-34){8}
    \DashLine(201,-34)(269,0){8}
    \Photon(201,-34)(269,-68){3.5}{5}
    \Vertex(133,-34){4}
    \Vertex(201,-34){4}

    \SetOffset(380,180)
    \PhotonArc(167,-34)(34.0,0,180){3.5}{8}
    \PhotonArc(167,-34)(34.0,180,360){3.5}{8}
    \DashLine(65,-34)(133,-34){8}
    \DashLine(201,-34)(269,0){8}
    \Photon(201,-34)(269,-68){3.5}{5}
    \Vertex(133,-34){4}
    \Vertex(201,-34){4}


    \SetOffset(-60,100)
    \DashLine(65,-34)(133,-34){8}
    \Photon(201,0)(269,0){3.5}{5}
    \Photon(201,-68)(269,-68){3.5}{5}
    \DashLine (133,-34)(201,0){8}
    \DashLine (201,0)(201,-68){8}
    \Photon (133,-34)(201,-68){3.5}{5}
    \Vertex(133,-34){4}
    \Vertex(201,0){4}
    \Vertex(201,-68){4}

    \SetOffset(50,100)
    \DashLine(65,-34)(133,-34){8}
    \Photon(201,0)(269,0){3.5}{5}
    \Photon(201,-68)(269,-68){3.5}{5}
    \DashLine (133,-34)(201,0){8}
    \Photon (201,0)(201,-68){3.5}{5}
    \Photon (133,-34)(201,-68){3.5}{5}
    \Vertex(133,-34){4}
    \Vertex(201,0){4}
    \Vertex(201,-68){4}

    \SetOffset(160,100)
    \DashLine(65,-34)(133,-34){8}
    \Photon(201,0)(269,0){3.5}{5}
    \Photon(201,-68)(269,-68){3.5}{5}
    \Photon (133,-34)(201,0){3.5}{5}
    \Photon (201,0)(201,-68){3.5}{5}
    \Photon (133,-34)(201,-68){3.5}{5}
    \Vertex(133,-34){4}
    \Vertex(201,0){4}
    \Vertex(201,-68){4}

    \SetOffset(270,100)
    \DashLine(65,-34)(133,-34){8}
    \Photon(201,0)(269,0){3.5}{5}
    \Photon(201,-68)(269,-68){3.5}{5}
    \ArrowLine (133,-34)(201,0)
    \ArrowLine (201,0)(201,-68)
    \ArrowLine (201,-68)(133,-34)
    \Vertex(133,-34){4}
    \Vertex(201,0){4}
    \Vertex(201,-68){4}

    \SetOffset(380,100)
    \PhotonArc(167,-34)(34.0,0,180){3.5}{8}
    \PhotonArc(167,-34)(34.0,180,360){3.5}{8}
    \DashLine(65,-34)(133,-34){8}
    \Photon(201,-34)(269,0){3.5}{5}
    \Photon(201,-34)(269,-68){3.5}{5}
    \Vertex(133,-34){4}
    \Vertex(201,-34){4}


    \SetOffset(50,20)
    \Photon(65,-34)(133,-34){3.5}{5}
    \Photon(201,0)(269,0){3.5}{5}
    \Photon(201,-68)(269,-68){3.5}{5}
    \Photon (133,-34)(201,0){3.5}{5}
    \Photon (201,0)(201,-68){3.5}{5}
    \Photon (133,-34)(201,-68){3.5}{5}
    \Vertex(133,-34){4}
    \Vertex(201,0){4}
    \Vertex(201,-68){4}

    \SetOffset(160,20)
    \Photon(65,-34)(133,-34){3.5}{5}
    \Photon(201,0)(269,0){3.5}{5}
    \Photon(201,-68)(269,-68){3.5}{5}
    \ArrowLine (133,-34)(201,0)
    \ArrowLine (201,0)(201,-68)
    \ArrowLine (201,-68)(133,-34)
    \Vertex(133,-34){4}
    \Vertex(201,0){4}
    \Vertex(201,-68){4}

    \SetOffset(270,20)
    \PhotonArc(167,-34)(34.0,0,180){3.5}{8}
    \PhotonArc(167,-34)(34.0,180,360){3.5}{8}
    \Photon(65,-34)(133,-34){3.5}{5}
    \Photon(201,-34)(269,0){3.5}{5}
    \Photon(201,-34)(269,-68){3.5}{5}
    \Vertex(133,-34){4}
    \Vertex(201,-34){4}

  \end{picture}
\end{center}
\caption{\label{figura4} Non null contributions 
to the {\texttt{sff}}, {\texttt{vff}}, 
{\texttt{sss}}, {\texttt{vss}}, {\texttt{svv}} and {\texttt{vvv}} 
${\mathrm R_2}$ effective 
vertices in the generalized $R_\xi$ gauges, with generic finite
$\xi$, $\xi_Z$, $\xi_A$. 
The corresponding analytical formulas associated to selected
topologies of
each generic
diagram are included in the files $\texttt{xxxxgentop.h}$, 
with $\texttt{xxxx}$ = $\texttt{sff}$, $\texttt{vff}$, 
$\texttt{sss}$, $\texttt{vss}$, $\texttt{svv}$ and
$\texttt{vvv}$,
respectively. 
For all diagrams including a $\texttt{vvvv}$ vertex see the comments
in Fig.~\protect\ref{figura2}.
}
\end{figure}
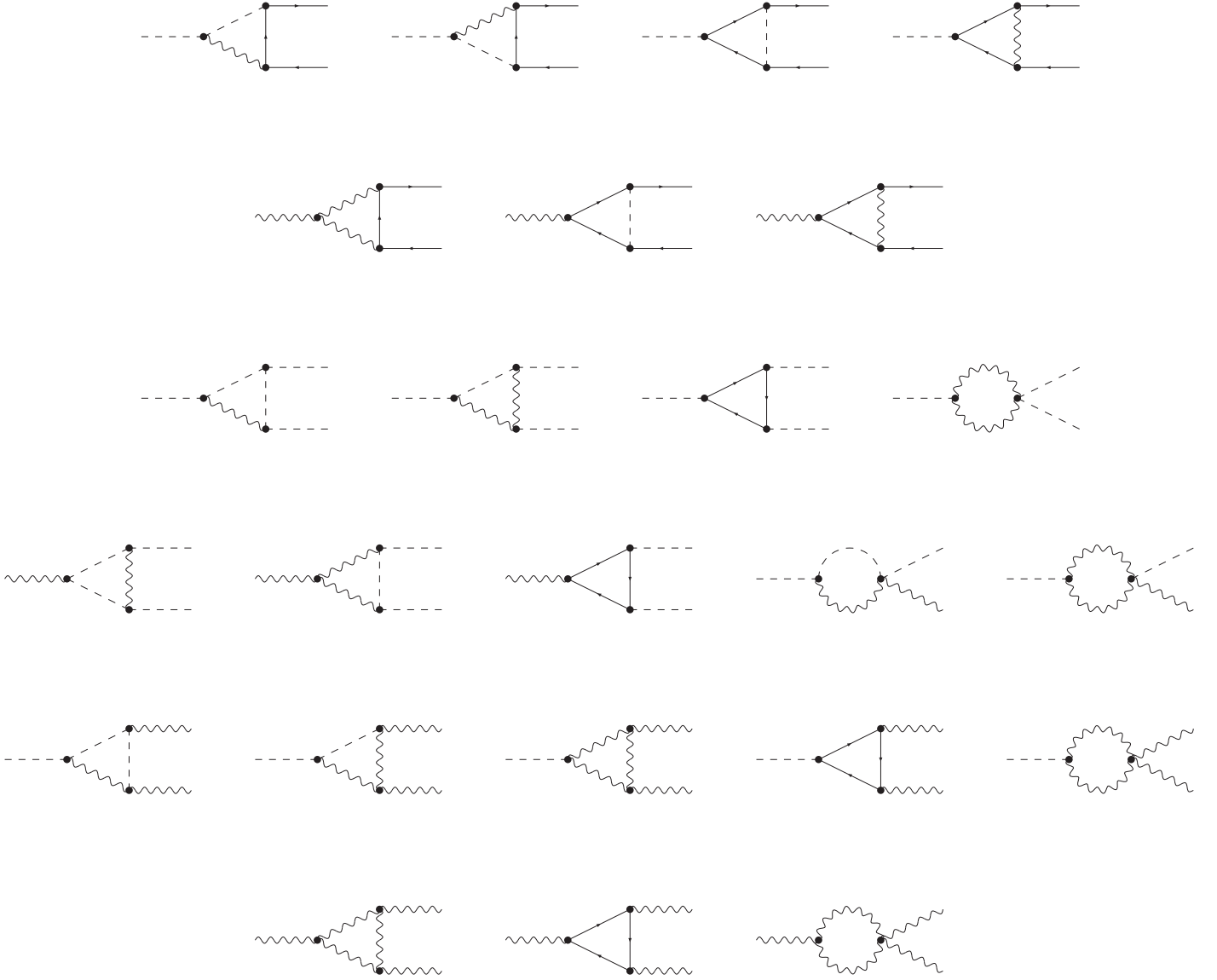

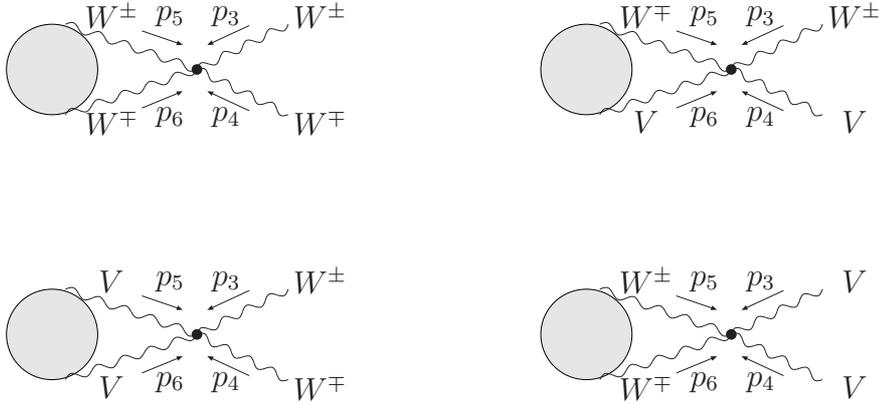
\begin{figure}[h!]
\begin{center}
  \begin{picture}(550,200) 
    \SetOffset(0,142)
    \SetScale{0.5}
    \SetWidth{0.5}
    \SetColor{Black}

    \SetOffset(0,142)

    \GCirc(123,-34){34.0}{0.9}

    \Photon(231,-34)(299,0){3.}{4}
    \Photon(231,-34)(299,-68){3.}{4}
    \Photon(133,0)(231,-34){3.}{5}
    \Photon(133,-68)(231,-34){3.}{5}

    \Vertex(231,-34){4.0}

    \LongArrow(190,-63)(219,-51)  
    \Text(106,-35)[]{{\Black{$p_6$}}}

    \LongArrow(190,-5)(219,-15)  
    \Text(106,3)[]{{\Black{$p_5$}}}

    \LongArrow(270,-1)(240,-15)  
    \Text(127,3)[]{{\Black{$p_3$}}}

    \LongArrow(270,-65)(240,-51)  
    \Text(127,-35)[]{{\Black{$p_4$}}}

    \Text(162,3)[]{$W^\pm$}
    \Text(162,-37)[]{$W^\mp$}
    \Text(84,3)[]{$W^\pm$}
    \Text(84,-37)[]{$W^\mp$}

    \SetOffset(200,142)

    \GCirc(123,-34){34.0}{0.9}

    \Photon(231,-34)(299,0){3.}{4}
    \Photon(231,-34)(299,-68){3.}{4}
    \Photon(133,0)(231,-34){3.}{5}
    \Photon(133,-68)(231,-34){3.}{5}

    \Vertex(231,-34){4.0}

    \LongArrow(190,-63)(219,-51)  
    \Text(106,-35)[]{{\Black{$p_6$}}}

    \LongArrow(190,-5)(219,-15)  
    \Text(106,3)[]{{\Black{$p_5$}}}

    \LongArrow(270,-1)(240,-15)  
    \Text(127,3)[]{{\Black{$p_3$}}}

    \LongArrow(270,-65)(240,-51)  
    \Text(127,-35)[]{{\Black{$p_4$}}}

    \Text(162,3)[]{$W^\pm$}
    \Text(162,-37)[]{$V$}
    \Text(84,3)[]{$W^\mp$}
    \Text(84,-37)[]{$V$}

    \SetOffset(0,42)

    \GCirc(123,-34){34.0}{0.9}

    \Photon(231,-34)(299,0){3.}{4}
    \Photon(231,-34)(299,-68){3.}{4}
    \Photon (133,0)(231,-34){3.}{5}
    \Photon (133,-68)(231,-34){3.}{5}

    \Vertex(231,-34){4.0}

    \LongArrow(190,-63)(219,-51)  
    \Text(106,-35)[]{{\Black{$p_6$}}}

    \LongArrow(190,-5)(219,-15)  
    \Text(106,3)[]{{\Black{$p_5$}}}

    \LongArrow(270,-1)(240,-15)  
    \Text(127,3)[]{{\Black{$p_3$}}}

    \LongArrow(270,-65)(240,-51)  
    \Text(127,-35)[]{{\Black{$p_4$}}}

    \Text(162,3)[]{$W^\pm$}
    \Text(162,-37)[]{$W^\mp$}
    \Text(84,3)[]{$V$}
    \Text(84,-37)[]{$V$}

    \SetOffset(200,42)

    \GCirc(123,-34){34.0}{0.9}

    \Photon(231,-34)(299,0){3.}{4}
    \Photon(231,-34)(299,-68){3.}{4}
    \Photon(133,0)(231,-34){3.}{5}
    \Photon(133,-68)(231,-34){3.}{5}

    \Vertex(231,-34){4.0}

    \LongArrow(190,-63)(219,-51)  
    \Text(106,-35)[]{{\Black{$p_6$}}}

    \LongArrow(190,-5)(219,-15)  
    \Text(106,3)[]{{\Black{$p_5$}}}

    \LongArrow(270,-1)(240,-15)  
    \Text(127,3)[]{{\Black{$p_3$}}}

    \LongArrow(270,-65)(240,-51)  
    \Text(127,-35)[]{{\Black{$p_4$}}}

    \Text(162,3)[]{$V$}
    \Text(162,-37)[]{$V$}
    \Text(84,3)[]{$W^\pm$}
    \Text(84,-37)[]{$W^\mp$}

  \end{picture}

\end{center}
\caption{\label{figura2} Diagrams in terms of generic
fields, including a {\texttt{vvvv}}
vertex, considered separately  
in our work. The charged vector bosons are
denoted by $W^\pm$ whereas the neutral vector bosons are denoted by $V$.
The first two diagrams are grouped together since they give rise
to the same $R_2$ formulas  
in the ${\texttt{xxvvgentop.h}}$ files, 
as well as the second two.}
\end{figure}

Furthermore, in case of fermionic loops, 
just for practical purposes,
we explicitly distinguish 
two generic diagrams corresponding to the same
topology, obtained one from the other at a fixed 
configuration of external legs, by simply changing the 
direction of the fermion flow 
(clockwise in the $\texttt{ff}$, $\texttt{fff}$, $\texttt{ffff}$ diagrams, 
and anticlockwise
in the $\texttt{ffac}$, $\texttt{fffac}$, $\texttt{ffffac}$ diagrams).  
In case of the $\texttt{ssvv}$ 
diagrams, an additional couple of topologies 
(clockwise and anticlockwise) appears, 
corresponding to the exchange of the second 
scalar with the first vector, that has to be considered separately
(see Fig.~\ref{figura5}). 

The symmetry factors 
have not been included at this level of the calculation.
They have instead been restored at the following step, 
i.e. when replacing generic fields with specific particles, 
accomplished
by means of do-loop procedures as explained below. 

All global variables, functions and constants used in the package
are declared in the file $\texttt{variables.h}$, including some explanations
of their meaning. In particular all SM particles, with their mass,
charge and isospin, are listed (see also Section~\ref{notation} 
on our notations). 

The expressions of all EW vertices, taken from Ref.~\cite{denner}, are
encoded in a specific file, named {\texttt{fillvertices.h}}, organized
in procedures. Each procedure dresses (i.e. fills) a particular
generic vertex (e.g. {\texttt{svv}}) with real specific particles.    
Special care has been taken in the treatment 
of the $\texttt{vss}$, $\texttt{vvv}$ and $\texttt{vvvv}$ vertices.
In fact, these vertices are not symmetric under the exchange of two generic
particles, so one has to distinguish different possibilities. 
In case of generic diagrams involving $\texttt{vvvv}$ vertices, we 
explicitly obtain different $\rm R_2$ formulas according to the nature (charged
$W$ or neutral $V$) of the four vectors entering the vertex, and
to their position in the diagram as external or internal particles. 
In general, one has to distinguish 4 cases corresponding to
the same topology, as shown in Fig.~\ref{figura2}: the case of
an internal ($W^+$, $W^-$) and an external ($W^+$, $W^-$) couple, the one of 
an internal ($W$, $V$) and an external ($W$, $V$) couple, the one of 
an internal ($V$, $V$) and an external ($W^+$, $W^-$) couple, and the one of
an internal ($W^+$, $W^-$) and an external ($V$, $V$) couple. 
In practice, the first two
cases and the second two cases can be grouped together, 
corresponding to the same formulas. 
We explicitly distinguish these possibilities in the formulas presented
in the files. A further configuration has also to be taken into account
in building the $W^+W^-W^+W^-$ $R_2$ effective vertex, 
corresponding to an internal ($W^+$, $W^+$) and an external ($W^-$, $W^-$) couple or viceversa.  

The procedure to fill generic masses with
the ones of real particles and some procedures that relate
particles with their charge conjugated, just for internal use,
are also included 
in the $\texttt{fillvertices.h}$ file.   
All procedures included in the $\texttt{fillvertices.h}$ file are valid in the
't Hooft Feynman gauge. The additional multiplicative $\xi$, $\xi_Z$ factors
appearing in some of the expressions of particle masses 
when 
considering the generalized $R_\xi$ gauges instead of the 't Hooft Feynman one, 
are always already included directly into the $\rm R_2$
formulas associated to the generic diagrams
listed in the $\texttt{xxxxgentop.h}$ files. In particular, 
we introduced
a $\texttt{csif}$ function, which gives the $\xi$ parameter
associated to each particle under consideration. To control 
its values the user is allowed to
modify the procedure $\texttt{fillcsi}$ in the $\texttt{fillvertices.h}$ 
file.   
Further warnings about the validity and the applicability 
of some of these formulas are explicitly mentioned in the files.  

The generic diagrams are then dressed by means of specific particles,
in such a way to sum together all 1-loop contributions corresponding to 
the same real external particles, by considering all different 
non-equivalent topologies.
The ${\rm R_2}$ effective vertices are built this way, 
by using the do-loop procedures included in the  
$\texttt{xxxxfeynrules.frm}$ files, with {\texttt{xxxx}}= {\texttt{ss}}, {\texttt{vs}}, {\texttt{vv}}, {\texttt{ff}}, {\texttt{sff}}, {\texttt{vff}}, {\texttt{sss}}, {\texttt{vss}}, {\texttt{svv}}, {\texttt{vvv}}, {\texttt{ssss}}, {\texttt{ssvv}}, {\texttt{vvvv}}. 

These do-loop procedures have the
advantage of leading to all results for each class of 
effective vertices joining specific particles
in an automatic way, without the need
of considering each effective vertex as a particular case, to be 
treated separately from the others according to the specific
external particles attached, and of
explicitly drawing all Feynman particle diagrams contributing to it. 
The most external do-loops run on the external particles, whereas
the other ones run on the internal particles. All possible combinations
of internal and external particles are considered at the do-loop level, 
then, the ones that do not exist in nature, are simply discarded 
due to the lack of the corresponding vertices in the $\texttt{fillvertices.h}$ 
procedures. The do-loop procedures have been tested against a completely
independent filling procedure, that works effective vertex by effective vertex, 
merely consisting in hand writing down, one by one, all diagrams contributing
to each effective vertex, on the basis of the output of a 
FeynArts~\cite{feynarts} computation, and summing the corresponding
$\rm R_2$ contributions together.
 
By simply changing the values of some control variables, 
the user has the possibility to modify the
output, e.g. to study the impact of different contributions
to each effective vertex joining specific particles, 
by selecting the box, triangle or
bubble diagram contributions only, or the fermionic loop contributions
only, or to choose among different gauges in the $R_\xi$ class, in particular 
the 't Hooft Feynman and 
the Landau gauge, or among different dimensional regularization schemes,
i.e. the Four Dimensional Helicity (FDH) scheme, 
or the 't Hooft Veltman (HV) one (different schemes differ for the 
treatment of the polarization vectors of the particles: 
in the FDH scheme all helicities are treated
in 4 dimensions, 
in the
HV scheme unobserved polarization vectors are continued
to 
$d$
dim, whereas the observed ones are kept in 4 dim~\cite{kos92}).

The particular cases here
mentioned can be obtained by simply changing the values of specific
variables included at the end of each do-loop procedure, as suggested
in Table~\ref{tabella1}, which shows the already implemented options. 

\begin{table}[h!]
\begin{center}
\begin{tabular}{|c|c|}
\hline
\multicolumn{2}{|c|}{choice of the dimensional regularization scheme} \\
\hline
{\it parameter} & {\it option} \\
\hline
{\texttt{lambdahv generic}} & default \\
{\texttt{lambdahv = 0}} & FDH scheme \\
{\texttt{lambdahv = 1}} & HV 
scheme \\
\hline
\end{tabular}
\vskip 0.8cm
\begin{tabular}{|c|c|}
\hline
\multicolumn{2}{|c|}{choice of the gauge} \\
\hline
{\it parameter} & {\it option} \\
\hline
{\texttt{csia, csiz, csi generic}} & generalized $R_\xi$ gauges, default \\
{\texttt{csia = csiz = csi}} & 1-parameter $R_\xi$ gauges \\
{\texttt{csia = csiz = csi = 1}} 
& 't Hooft-Feynman gauge \\
\hline
\end{tabular}
\vskip 0.8cm
\begin{tabular}{|c|c|c|c|c|}
\hline
\multicolumn{5}{|c|}{selection of different $\rm R_2$ contributions} \\
\hline
\multicolumn{4}{|c|}{$parameter$} & \multirow{2}{*}{$option$} \\
\cline{1-4}
\texttt{dummyb} & \texttt{dummyc} & \texttt{dummyd} & \texttt{dummyf} & 
$\,$ \\
\hline
\multirow{2}{*}{1} & \multirow{2}{*}{1} & 
\multirow{2}{*}{1} & \multirow{2}{*}{1} & 
default, all contributions (boxes, triangles, bubbles)  \\
$\, $ & $\, $ & $\, $ & $\, $ & to each effective vertex are kept \\
\hline 
\multirow{2}{*}{1} & \multirow{2}{*}{0} & \multirow{2}{*}{0} 
& \multirow{2}{*}{0} &
only bubble contributions are selected \\ 
$\, $ & $\, $ & $\, $ & $\, $ & (excluding bubble fermionic loops, if any)\\
\hline
\multirow{2}{*}{0} & \multirow{2}{*}{1} & 
\multirow{2}{*}{0} & \multirow{2}{*}{0} &
only triangle contributions are selected \\  
$\, $ & $\, $ & $\, $ & $\, $ & (excluding fermionic triangles, if any)\\
\hline
\multirow{2}{*}{0} & \multirow{2}{*}{0} & 
\multirow{2}{*}{1} & \multirow{2}{*}{0} & 
only boxes contributions are selected \\ 
$\, $ & $\, $ & $\, $ & $\, $ & (excluding fermionic boxes, if any)\\
\hline
0 & 0 
& 0 & 1 &
only fermionic-loop contributions are selected \\
\hline
1 & 1 & 1 & 0 & only non-fermionic-loop contributions are selected \\
\hline
\end{tabular}
\caption{\label{tabella1} Options already implemented for the 
control variables the user is allowed to tune at the end of the
do-loop procedures in the
$\texttt{xxxxfeynrul.frm}$ files, to select a particular gauge 
and regularization scheme,
or to enlighten particular contributions to the $R_2$ effective
vertices coming from specific set of diagrams. 
} 
\end{center}
\end{table}

The fermionic contribution is gauge independent, i.e. it is the same in all
gauges, and this is the reason why we allow its selection separately
from the other contributions.
The part of the results proportional to $\lambda_{HV}$ is also gauge 
independent, i.e. the choice of a regularization scheme can be safely
performed in an independent way with respect to the gauge choice.

The package is available as a zipped tar archive at the web address\\
{\texttt{http://www.ugr.es/$\sim$garzelli/R2SM}. 
After decompressing it, all files with a $\texttt{.frm}$ extension can be runned 
by using a working installation of FORM~\cite{form}, 
downloadable from the web at the address 
{\texttt{http://www.nikhef.nl/$\sim$form}, that needs to be preinstalled 
by the user. In particular, we extensively 
used the version 3.2 (May 2008) of the 
FORM package. We verified that the same output is produced when running
the code with FORM version 3.3 (August 2010), now available on the web.

\section{A check of the gauge invariance of the R
contribution to (renormalized) S-matrix elements: 
the H~$\rightarrow$~$\gamma\gamma$ decay}
\label{hgg}

The Green functions in general depend on the gauge choice, but 
renormalized S-matrix elements must be gauge independent, since
they correspond to physical observable quantities. 

The contributions to renormalized S-matrix elements of the CC terms
and of the R terms are separately gauge invariant. 
As a particular check of this point and of the correctness
of our calculations, we have explicitly proven that the total R 
contribution, ${\rm R_1 + R_2}$,  
to the S-matrix element corresponding to the  
$H \rightarrow \gamma \gamma$ physical decay
process at 1-loop is gauge independent.
The code we have written to produce this result is included
in the subdirectory $\texttt{RtotHAA}$ of our package. 
All new variables specific to $\rm R_1$ calculations, are included
in the $\texttt{additionalvariables.h}$ file.
The non trivial ${\rm R_1}$ contributions 
to 1-loop generic diagrams 
including as external particles a generic scalar and two generic 
vectors, are presented in the $\texttt{svvR1gentop.h}$ file.
The calculation of the ${\rm R_1}$ terms has been performed in a similar way to 
the one of the ${\rm R_2}$ terms, 
the main difference being the fact that only the 
4-dim part of the numerator $\bar{N}(\bar{q})$ can contribute to ${\rm R_1}$. 
The ${\rm R_1}$ part of the tensor integrals involved
has been  
extracted by applying the Passarino-Veltman reduction technique~\cite{pasve}
in a straightforward way,
by disregarding the contribution of the scalar functions $A$, $B$, $C$, $D$ 
which, by definition, already contribute to the CC part of the amplitude.
${\rm R_1}$ effective vertices corresponding to specific 
$\texttt{svv}$ configurations 
involving physical external particles, can be obtained by dressing the generic
diagrams mentioned above with specific particles, 
thanks to a do-loop procedure, and by conveniently summing the
corresponding contributions together,
in a way analogous to the one followed 
to build the ${\rm R_2}$ effective
vertices. The do-loop procedure we have used to calculate the R = 
${\rm R_1}$ + ${\rm R_2}$ effective vertices for all specific {\texttt{svv}} 
combinations 
is presented in the file $\texttt{svvRfeynrules.frm}$.
Finally, in the file $\texttt{haa.frm}$, 
we show that the R contribution to the S-matrix element of the 
$H\gamma\gamma$ decay process
is gauge invariant,
i.e. the result is independent of any gauge parameter 
(and it remains the same even in
the limit $\xi, \, \xi_Z \rightarrow \infty$ and $\xi_A \rightarrow 1$: 
we have repeated this calculation in the unitary gauge, making
these limits before the integration, and
obtained the same final result for the R part of the S-matrix element,
also in agreement with Ref.~\cite{pasbook}). 
  
\section{Conclusions}
\label{conclusions}

We have presented an analytical package, written in FORM and  
available on the web, to calculate the ${\rm R_2}$ effective Feynman rules 
by means of which one can compute the ${\rm R_2}$ contribution to 
1-loop amplitudes for whichever process in the SM of the EW 
interactions. As recommended in the literature, 
we have chosen to work in the $R_\xi$ class of gauges,
identified by finite values of the $\xi$, $\xi_Z$, 
$\xi_A$ parameters, 
due to the good suitability of this class to loop calculations,
since 
the UV behaviour of the theory is not spoiled by the particular 
expression of the propagators of the massive gauge bosons
in this class.
We have explicitly verified that this general recommendation also apply to 
the calculation of the ${\rm R_2}$ contribution alone, to 1-loop amplitudes: 
even if, from the technical point of view, analytical calculations 
of the R part of the amplitude can be performed as well in 
the unitary gauge, the $R_\xi$ gauges have 
by far better properties in terms of the compactness and the simplicity 
of the final ${\rm R_2}$ analytical formulas with respect to the 
unitary gauge.  

We have also offered a check of the gauge-invariance of the total 
R contribution, ${\rm R_1}$~+~${\rm R_2}$, to the renormalized 
S-matrix element for the $H\gamma\gamma$ process. 

The package is modular, and different pieces can be conveniently
reintegrated in other calculations. 
The user is allowed to play with some parameters, in order to specify
a particular gauge in the $R_\xi$ class or a particular dimensional
regularization scheme, and to enlighten different partial 
contribution to the ${\rm R_2}$ effective vertices. These options allow 
everybody to reproduce, as particular cases, some of the results 
presented in our previous  papers~\cite{ultimo,ewrational}, in a 
straightforward way.

Finally, we think that our effort can be considered a first seed
towards the more complex task of automatically recovering, given a 
Lagrangian for whichever model of particle interactions, all
${\rm R_2}$ effective Feynman Rules for the theory at hand.

\section{Acknowledgments}
We are grateful to R. Pittau and R. Kleiss for many useful discussions and
suggestions. The work of M.~V.~G. was supported by the italian INFN, 
the 
work of I.~M. was supported by the RTN European 
Programme MRTN-CT-2006-035505 (HEPTOOLS, Tools and Precision 
Calculations for Physics Discoveries at Colliders). 
M.~V.~G. also acknowledges her partecipation in the 
MEC Project FPA2008-02984, thanks to
which a brief stay of Y.~M. at the University of Granada, 
crucial for developing this research together, 
was possible. 

\section{Appendix: Notation}
\label{notation}

We closely follow the conventions in the paper~\cite{denner}.
We adopt the Bjorken-Drell or ``mostly minus'' metric convention
$\eta_{\mu \nu} = diag(+1,-1,-1,-1)$. 

We consider 1-loop amplitudes with at most 4 external legs.
The momenta of the external particles are denoted by $p_1$, $p_2$, $p_3$, $p_4$
and are all supposed to be incoming.
In the 2 point effective vertices, the Lorentz indices associated to the 
external bosons whose momenta are $p_1$ and $p_2(=-p_1)$, 
are respectively denoted by $\alpha$ and $\delta$.
In the 3 point effective vertices, the Lorentz indices associated to the 
external bosons whose momenta are $p_1(=-p2-p3)$, $p_2$ and $p_3$, 
are denoted by $\tau$, $\omega$ and $\chi$, respectively.
Finally, in the 4 point effective vertices, the Lorentz indices 
associated to the 
external bosons whose momenta are $p_1$, $p_2$, $p_3$ and $p_4(=-p1-p2-p3)$ 
are denoted by $\alpha$, $\beta$, $\tau$ and $\chi$.

We worked in the generalized $R_\xi$ gauge, by allowing 3 different
parameters $\xi_A$, $\xi_Z$ and $\xi$ (\texttt{csia}, \texttt{csiz}
and \texttt{csi} in our files) in the gauge fixing term
(usually taken to be equal in the standard $R_\xi$ gauge). 
The $\lambda_{HV}$ ($\texttt{lambdahv}$ in our files) 
dependence, identifying the dimensional regularization scheme, 
has been made explicit in all results. 
We checked that the contribution
proportional to $\lambda_{HV}$ for each effective vertex is always invariant 
with respect to gauge transformations. 

\begin{table}
\begin{center}
\begin{tabular}{|c|c|c|}
\hline
\multicolumn{3}{|c|}{Particle content of the EW sector of the SM} \\
\hline
$generic$ $particle$ & $specific$ $particles$ & $notation$ $in$ $our$ FORM $files$ \\
\hline
scalars $s$ & $H$, $\chi$, $\phi^+$, $\phi^-$ &
\texttt{hsca, chisca, phiplussca, phimensca} \\
\hline
vectors $v$ & $A$, $Z$, $W^+$, $W^-$ &
{\texttt{avec, zvec, wplusvec, wmenvec}} \\
\hline
\multirow{2}{*}{ghosts $fp$} & $u_A$, $u_Z$, $u_W^+$, $u_W^-$ &
{\texttt{gavec, gzvec, gwplusvec, gwmenvec}} \\
$\, $
& $\bar{u}_A$, $\bar{u}_Z$, $\bar{u}_W^+$, $\bar{u}_W^-$ &
{\texttt{gavecbar, gzvecbar, gwplusvecbar, gwmenvecbar}} \\  
\hline
\multirow{2}{*}{charged leptons $f$} & $e$, $\mu$, $\tau$ & {\texttt{ele, muele, tauele}} \\
$\, $
& $\bar{e}$, $\bar{\mu}$, $\bar{\tau}$ & {\texttt{elebar, muelebar, tauelebar}}\\
\hline
\multirow{2}{*}{neutral leptons $f$} & $\nu_e$, $\nu_\mu$, $\nu_\tau$ & {\texttt{nuele, numuele, nutauele}} \\
$\, $
& $\bar{\nu}_e$, $\bar{\nu}_\mu$, $\bar{\nu}_\tau$ & {\texttt{nuelebar, numuelebar, nutauelebar}}\\
\hline
\multirow{2}{*}{up-type quarks $f$} &  $u$, $c$, $t$ & {\texttt{uquark, cquark, tquark}} \\
$\, $
& $\bar{u}$ $\bar{c}$, $\bar{t}$ & {\texttt{ubarquark, cbarquark, tbarquark}}\\
\hline
\multirow{2}{*}{down-type quarks $f$} & $d$, $s$, $b$ & {\texttt{dquark, squark, bquark}} \\
$\, $
& $\bar{d}$ $\bar{s}$, $\bar{b}$ & {\texttt{dbarquark, sbarquark, bbarquark}}\\
\hline
\end{tabular}
\end{center}
\caption{\label{tabella2} Notation used in this work for the particle content of the EW sector of the SM. The Faddeev-Popov-De Witt ghosts are only relevant for the calculation of
the $\rm R_1$ terms, whereas all other particles enter in the calculation
of both the $\rm R_2$ and the $\rm R_1$ contributions to 1-loop amplitudes.}
\end{table}

In our FORM files, we labelled the particles of the EW sector of the SM 
according to the notation
of Table~\ref{tabella2}. 
Actually, we limited to consider only
one weak-isospin family of quarks and one weak-isospin family of leptons.
Anyway, the results can be extended to three families 
in a straightforward way, 
by just working on the fermionic
part of the do-loop procedures in all files with a $\texttt{.frm}$ extension. 
The number of quark colors can be fixed by the user by a proper
choice of the value of the variable {\texttt{ncol}} at the end
of each do-loop procedure. 
The contribution of fermionic loops is independent
of the gauge choice, since all propagators and coupling constants
involving fermions are the same in all gauges. Thus, for the parts of
our final results including fermion loops, and for their generalization to
three weak-isospin families, 
the reader can safely refer to the analytical formulas already presented 
in our previous paper~\cite{ewrational}.

The mass of each SM particle is denoted by adding an ``$\texttt{m}$'' before
the name of the particle (e.g. {\texttt{mhsca}} is the mass of the Higgs scalar,
{\texttt{mwvec}} is the mass of the $W$ gauge boson).
The pole masses of the unphysical scalars $\chi$ and $\phi^\pm$ are 
reduced to the masses of the corresponding massive 
gauge bosons by $m_\chi =\sqrt{\xi_Z} m_Z$ and $m_\phi = \sqrt{\xi} m_W$. 
We also widely apply
the relation $m_Z = m_W/\mathrm{cos} \theta_W$, and thus express our
results in terms of $m_W$. The sine and cosine of the Weinberg angle 
are denoted by {\texttt{sinwei}} and {\texttt{coswei}}, respectively, 
and the $e$ coupling constant is denoted by $\texttt{eem}$.  

The third component of the weak-isospin of each fermion is denoted
by including the prefix ``$\texttt{i3}$'' before the name of the particle:
$\texttt{i3nuele}$ (= 1/2) corresponds to the $\nu_e$, 
$\texttt{i3ele}$ (= -1/2) corresponds to the $e^-$, 
$\texttt{i3u}$ (= 1/2) and $\texttt{i3d}$ (= -1/2) correspond to the 
$u$ and $d$ quark quantum numbers, respectively. 

The projector operators $\Omega_+ = \frac{1 + \gamma_5}{2}$ and
$\Omega_- = \frac{1 - \gamma_5}{2}$ are denoted by {\texttt{omegaplus}}
and {\texttt{omegaminus}}, respectively. 
As for the treatment of $\gamma_5$, the reader can refer to the
comments already presented in Ref.~\cite{ewrational}.

\end{document}